\documentclass[a4paper]{article}
\usepackage[dvipsnames]{xcolor}
\usepackage[utf8]{inputenc}
\usepackage[english]{babel}
\usepackage{amsmath}
\usepackage{amssymb}
\usepackage{amsthm}

\usepackage{hyperref}
\usepackage{multicol}
\usepackage{tikz}
\usetikzlibrary{shapes,arrows}
\usepackage[nottoc]{tocbibind}
\usepackage{graphicx}
\usepackage{bbold}
\usepackage{csquotes}
\usepackage{comment}
\usepackage{sidecap}

\usepackage[a4paper,width=150mm,top=25mm,bottom=45mm,bindingoffset=6mm]{geometry}
\usepackage{fancyhdr}
\usepackage{mathrsfs}
\usepackage[normalem]{ulem}

\usepackage{import}

\definecolor{nred}{rgb}{0.7,0,0}
\definecolor{ngreen}{rgb}{0,0.6,0}

\usepackage[doi=false,isbn=false, url=false, date=year, backend=biber
,maxbibnames=99]{biblatex} 
\renewbibmacro*{journal+issuetitle}{%
  \usebibmacro{journal}%
  \setunit*{\addspace}%
  \iffieldundef{series}
    {}
    {\newunit
     \printfield{series}%
     \setunit{\addspace}}%
  \usebibmacro{volume+number+eid}%
  \setunit{\bibpagespunct}%
  \printfield{pages}%
  \setunit{\addspace}%
  \usebibmacro{issue+date}%
  \setunit{\addcolon\space}%
  \usebibmacro{issue}%
  \newunit}

\renewbibmacro*{note+pages}{%
  \printfield{note}%
  \newunit}

\DeclareMathOperator{\aut}{Aut}

\newcommand{\E}{\mathbb{E}}
\newcommand{\tr}{\mathrm{tr}}
\newcommand{\e}{\mathrm{e}}
\newcommand{\diam}{\mathrm{diam}}
\newcommand{\dd}{\mathrm{d}}
\renewcommand{\i}{\mathrm{i}}
\renewcommand{\Phi}{\varPhi}
\renewcommand{\Psi}{\varPsi}
\renewcommand{\Sigma}{\varSigma}
\newcommand{\epsi}{\varepsilon}

\newcommand{\g}{\gamma}
\newcommand{\La}{\Lambda}

\newcommand{\w}{\omega}
\newcommand{\N}{\mathbb{N}}
\newcommand{\Z}{\mathbb{Z}}
\newcommand{\C}{\mathbb{C}}

\newcommand{\mA}{\mathcal{A}}
\newcommand{\R}{\mathbb{R}}
\newcommand{\eps}{\varepsilon}
\newcommand{\ODeps}{^{\mathrm{OD} \varepsilon}   }
\newcommand{\OD}[1][]{   ^{\mathrm{OD}#1}   }
\newcommand{\DI}[1][]{   ^{\mathrm{D}#1}   }
\newcommand{\mL}{\mathcal{L}}

\usepackage{mathtools}

\DeclarePairedDelimiter\norm{\lVert}{\rVert}

\DeclarePairedDelimiter\br{\lparen}{\rparen}

\theoremstyle{plain}
\newtheorem{theorem}{Theorem}[section]
\newtheorem{lemma}[theorem]{Lemma}
\newtheorem{proposition}[theorem]{Proposition}

\theoremstyle{definition}
\newtheorem{definition}[theorem]{Definition}

\theoremstyle{remark}
\newtheorem{remark}[theorem]{Remark}

\newcommand\numberthis{\addtocounter{equation}{1}\tag{\theequation}}

\usepackage{parskip}

\usepackage{colonequals}

\usepackage{fullpage}

\title{A note on Hall conductance and Hall conductivity in interacting Fermion systems}
\author{Stefan Teufel
\texorpdfstring{\footnote{\parbox[t]{.7\textwidth}{
                \foreignlanguage{ngerman}{Fachbereich Mathematik,  Universität Tübingen,\\
                Auf~der~Morgenstelle~10, 72076~Tübingen,} Germany
            }
        }
    }{}%
\and Marius Wesle%
\texorpdfstring{%
        \footnotemark[1]
    }{}%
}
\date{\today}

\begin{document}

\maketitle

\begin{abstract}
In this note we consider lattice fermions on $\Z^2$ with a gapped ground state and show how to apply the NEASS approach to linear response to derive a formula for the \textit{Hall conductance} in terms of    the ground state  expectation of a commutator of modified step functions. This formula is usually derived by a charge pumping argument going back to Laughlin. Here we show that it can also be obtained as the linear response coefficient of the \textit{microscopic} current response to an adiabatic increase of the chemical potential on a half plane (or more generally on   any   cone-like region).  Indeed, in a manner reminiscent of the bulk-boundary correspondence, we show that raising the chemical potential in any cone-like region gives rise to a current that flows along its boundary and is nearly linear in the increase in chemical potential.  We also discuss the connection with the double commutator formula with modified position operators for the \textit{Hall conductivity} derived in \cite{wmmmt2024} as  the linear response coefficient of   the \textit{macroscopic} current response to the adiabatic application of a constant electric field.
\end{abstract}

\section{Introduction}

It is an important step in the theoretical understanding of the quantum Hall effect to show that the Hall current response of a gapped system in its ground state is nearly linear in the applied field. Mathematical proofs of this fact have been based on basically two different ways of modeling such systems and the applied field, namely the Laughlin pump approach and the NEASS approach. 
Roughly speaking, in the Laughlin pump approach, the spatial geometry of the system is assumed to be a torus and the driving is implemented by adiabatically threading a time-dependent magnetic flux through the torus. It is then shown that in one cycle of this charge pump an integer number of charges are transported over a fiducial circle up to corrections smaller than any power of the adiabatic parameter \cite{klein1990power,bachmann2021exactness}. The torus geometry has the advantage that one can model finite systems without boundaries. The elimination of boundaries is important, because otherwise, for non-trivial topological insulators, so-called edge states would close the spectral gap. The NEASS approach avoids boundaries by working directly on an infinitely extended system. In \cite{wmmmt2024} the driving is modeled by adiabatically applying a constant electric field over the whole space. It is then shown that the induced current density is   linear in the applied field up to corrections smaller than any power of the field strength.

In particular, the two approaches propose two different definitions of the relevant response coefficients. Not surprisingly, the resulting formulas for the response coefficients also differ slightly, but are expected to  lead to the same numerical values. 
This equality is well known for non-interacting systems (see \cite{mm22} and references therein) and has been shown under additional conditions in \cite{wmmmt2024} also for interacting systems. Both expressions have the form of a ground state expectation of a ``double commutator'', where in the charge pump approach one arrives at the ground state expectation of the commutator of two ``step functions'', while in the original NEASS approach one arrives at the ground state expectation per unit volume of the commutator of two position operators. This reflects the fact that in the charge pump approach the response is localized and independent of system size, we will say microscopic for short, whereas in the original  NEASS approach the response is a macroscopic homogeneous current density.

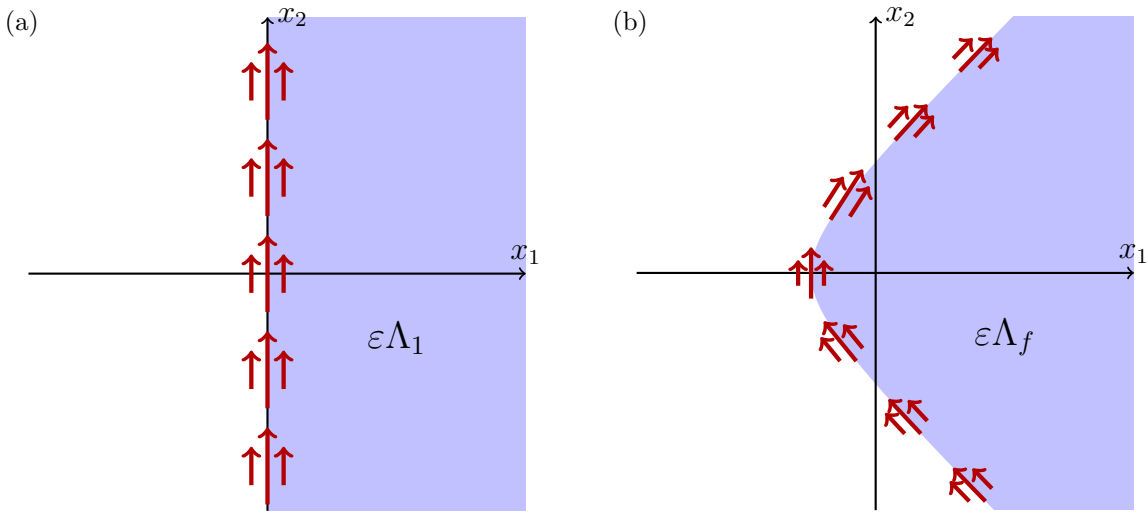
\begin{figure}[ht] \label{fig:current}
\usetikzlibrary{patterns} 

\begin{tikzpicture}[scale=.85,baseline=(current bounding box.south)]
  \begin{scope}
    \clip (0,-4) rectangle (4,4);
    \fill[color=blue!60,opacity=0.4] (-1,-3.7) rectangle (4,4);
  \end{scope}

  \draw[->,thick] (-3.7,0) -- (4,0) node[above] {\large$x_1$};
  \draw[->,thick] (0,-3.7) -- (0,4) node[right] {\large$x_2$};

    \draw[color=nred,->,ultra thick] (0,-.6) -- (0,.6);
    \draw[color=nred,->,ultra thick] (0,.9) -- (0,2.1);
    \draw[color=nred,->,ultra thick] (0,2.4) -- (0,3.6);
    \draw[color=nred,->,ultra thick] (0,-2.1) -- (0,-.9);
    \draw[color=nred,->,ultra thick] (0,-3.6) -- (0,-2.4);      
    
    \draw[color=nred,->,ultra thick] (-.25,-.3) -- (-.25,.3);
    \draw[color=nred,->,ultra thick] (.25,-.3) -- (.25,.3);
     \draw[color=nred,->,ultra thick] (-.25,1.2) -- (-.25,1.8);
    \draw[color=nred,->,ultra thick] (.25,1.2) -- (.25,1.8);
     \draw[color=nred,->,ultra thick] (-.25,-1.8) -- (-.25,-1.2);
    \draw[color=nred,->,ultra thick] (.25,-1.8) -- (.25,-1.2);
     \draw[color=nred,->,ultra thick] (-.25,-3.3) -- (-.25,-2.7);
    \draw[color=nred,->,ultra thick] (.25,-3.3) -- (.25,-2.7);
     \draw[color=nred,->,ultra thick] (-.25,2.7) -- (-.25,3.3);
    \draw[color=nred,->,ultra thick] (.25,2.7) -- (.25,3.3);

  \node at (2,-1) {\Large$ \eps  \Lambda_1$};
  \node at (-3.8,3.9) {(a)};
\end{tikzpicture}%
\qquad
\begin{tikzpicture}[scale=0.85, baseline=-97pt]
  \begin{scope}
    \clip 
    plot[smooth,domain=-1:3.5,samples=200] (\x, {sqrt((\x+2)*(\x+2)-1)})
      -- (4,4) -- (4,-4)
      -- plot[smooth,domain=3.5:-1,samples=200] (\x, {-sqrt((\x+2)*(\x+2)-1)})
      -- cycle;
    \fill[color=blue!60,opacity=0.4] (-3.7,-3.7) rectangle (4,4);
  \end{scope}
  \draw[->,thick] (-3.7,0) -- (4,0) node[above] {\large$x_1$};
  \draw[->,thick] (0,-3.7) -- (0,4) node[right] {\large$x_2$};

 \draw[->,nred,ultra thick] (-1,-0.4) -- ++(0,0.8);
\draw[->,nred,ultra thick] (-1.2,-0.2) -- ++(0,0.4);
\draw[->,nred,ultra thick] (-.8,-0.2) -- ++(0,0.4);
  
  \foreach \xv in {1.3,2.3,3.3} {
    \pgfmathsetmacro\yv{sqrt(\xv*\xv - 1)}
    \pgfmathsetmacro\m{(\xv/\yv)}
    \draw[->,nred,ultra thick]
      (\xv-2,\yv) -- ++(0.5, {0.5*\m});
      \draw[->,nred,ultra thick]
      (\xv-2.1,\yv+0.2) -- ++(0.3, {0.3*\m});
      \draw[->,nred,ultra thick]
      (\xv-1.7,\yv+0.05) -- ++(0.3, {0.3*\m});
  }

  \foreach \xv in {1.7,2.7,3.7} {
    \pgfmathsetmacro\yv{sqrt(\xv*\xv - 1)}
    \pgfmathsetmacro\m{(\xv/\yv)}
    \draw[->,nred,ultra thick]
      (\xv-2,-\yv) -- ++(-0.5, {0.5*\m});
      \draw[->,nred,ultra thick]
      (\xv-2.25,-\yv) -- ++(-0.3, {0.3*\m});
      \draw[->,nred,ultra thick]
      (\xv-1.9,-\yv+0.2) -- ++(-0.3, {0.3*\m});
  }
    \node at (2,-1) {\Large$ \eps  \Lambda_f$};
    \node at (-3.8,3.9) {(b)};
\end{tikzpicture}\vspace{-10mm}

\caption{
(a) In a topological insulator, an increase in the chemical potential in the right half-plane $\Lambda_1$ by~$\eps$ leads to an additional current flowing along the potential step (vertical axis),  although the spectral gap presumably does not close. The total current through the line $\{x_2=0\}$ can  formally be computed as the expectation of $\i[H_0,\Lambda_2] = \frac{\dd}{\dd t} \e^{\i H_0 t} \Lambda_2\e^{-\i H_0 t}|_{t=0}$, i.e.\ as the rate of change of the charge  in the upper half plane $\Lambda_2$, in the NEASS $\w_\eps$ (which is presumably the ground state of $H_\eps$). However, since in infinite volume this expectation, in general,  does not exist, one needs to subtract in \eqref{def:delta} the expectation of $\i[H_0,\Lambda_2]$ in the initial ground state $\w_0$. This difference is the additional current induced by the increase of the chemical potential and is, as we show, nearly linear in $\eps$. (b) The same is true when the chemical potential is increased on any   infinite cone-like set. Along the boundary of any such set, a current flows that is  nearly linear in $\eps$ and whose magnitude is the same for all such sets up to terms of order $\mathcal{O}(\eps^\infty)$.}
\end{figure}

In the present note we further elaborate the connections between the two approaches by applying the NEASS approach to the adiabatic switching of a potential step along an infinite line in an infinite system. The response of the system is an \textit{additional} current flowing along the potential step, additional with respect to such currents that may already be present in the initial gapped ground state, see also Figure~\ref{fig:current}. 
We argue that this is the natural generalization of the charge pump approach to infinitely extended systems, and we obtain the analog of the ``commutator of step function'' formula also in this setting up to corrections asymptotically smaller than any power of the height of the potential step.
While our results are not conceptually surprising, they do shed some light on the   relationship between the two approaches and also require technical improvements that we expect to be useful for future research. For example, the results of the present paper can be used to generalize the results of \cite{wmmmt2024} for the macroscopic current response of periodic systems to systems without any periodicity or homogeneity except for the spectral gap assumption \cite{mtw2025}.

Let us now describe our   results in some more detail.
We consider a system described by a short-range Hamiltonian $H_0$ on $\Z^2$ that initially starts in  a   gapped ground state $\omega_0$. Then we compute the response to the perturbation $\varepsilon \,\Lambda_1$,  where $\Lambda_1$ and $\Lambda_2$ are the number operators of the right and upper half-planes respectively. 
In fact, our results apply to a more general class of `step functions', which are not necessarily aligned with the lattice structure. However, for simplicity, in the discussion in the introduction we stick  to $\La_1$ and $\La_2$. 
It is shown in \cite{Teufel2020,HenheikTeufel2022,BTW2025} that adiabatically switching on this perturbation drives the system into a   state $\w_\eps$, which in general is a NEASS, but, for the specific perturbation considered here, presumably is   still a gapped ground state for the perturbed Hamiltonian $H_\varepsilon := H_0 + \varepsilon \Lambda_1$ and  $\eps$ small enough.

We are interested in the current response of the system, that is in the current flowing along  the potential step introduced by adding the perturbation $\eps\Lambda_1$. More precisely, we compute the   expectation value of the current flowing into the upper half-plane in the state $\omega_\varepsilon$ compared to $\omega_0$,
\begin{equation}\label{def:delta}
 \Delta J_\eps := \w_\eps(\i[H_0,\Lambda_2])-  \w_0(\i[H_0,\Lambda_2])\coloneq \sum_{x \in \Z^2}\big( \w_\eps(\i[H_0,\Lambda_2]_x) -\w_0(\i[H_0,\Lambda_2]_x)\big)\,,   
\end{equation}
where, at least formally,    $\i[H_0,\Lambda_2] = \frac{\dd}{\dd t} \e^{\i \mL_{H_0} t} \Lambda_2|_{t=0}$ is the time-derivative of the charge in the upper half-plane.
 Note that $\i[H_0,\Lambda_2]$ is not a quasi-local observable, but   a non-convergent sum $\sum_{x\in\Z^2}\i[H_0,\Lambda_2]_x$ of quasi-local terms $\i[H_0,\Lambda_2]_x\in \mA$ (see Section~\ref{sec:setup} for details),  and thus the expectations $\w_\eps(\i[H_0,\Lambda_2])$ and $\w_0(\i[H_0,\Lambda_2])$ are  apriori not well defined.  Indeed, while the sum on the right hand side of \eqref{def:delta}  converges absolutely, this is in general not the case for the individual sums. We discuss below why, in contrast to the current density response considered in \cite{wmmmt2024}  and to the finite volume situation with torus geometry considered in \cite{bachmann2021exactness}, in the present situation it is necessary to consider the current response relative to the current in the ground state.

Our main result, Theorem~\ref{thm:main}, states that the current response $\Delta J_\eps$ is nearly linear in $\eps$,
 \begin{align}\label{main_intro}
        \Delta J_{\eps} =  \eps \, \w_0( \i\,[\, \La_2\OD,\, \La_1\OD\,]) + \mathcal{O}(\eps^\infty)\, ,
    \end{align}
where the linear coefficient $ \w_0( \i\,[\, \La_2\OD,\, \La_1\OD\,])$, the Hall conductance, is given by the double-commutator formula for step functions. Here $\Lambda_j\OD$ denotes the off-diagonal part of $\Lambda_j$ with respect to $\omega_0$, see Definition~\ref{def:OD}, which is an interaction localized along the vertical axis for $\Lambda_1\OD$ resp.\ the horizontal axis for $\Lambda_2\OD$. Their commutator $[\, \La_2\OD,\, \La_1\OD\,]$ is a quasi-local operator localized near the origin and thus $\w_0( \i\,[\, \La_2\OD,\, \La_1\OD\,])$ is well defined.

Furthermore, we show that the Hall conductance $ \w_0(\i \, [\, \La_2\OD,\, \La_1\OD])$ 
\begin{itemize}
\item depends only on $\omega_0$ but not on $H_0$,  even though the off-diagonal map $\La_j\mapsto \La_j\OD$ does (Theorem~\ref{thm:main}),
\item  is insensitive to changes of the choice of origin, the angles of the half planes with respect to the lattice and each other, as well as a more general class of deformations of the two regions, as long as the relative orientation of the boundaries is preserved    (Theorem~\ref{thm:indep}),
\item is constant within gapped phases (Theorem~\ref{thm:const}),
\item and agrees with the Hall conductivity   $\overline \w_0(\i \, [\, X_2\OD,\, X_1\OD])$ derived for periodic systems in \cite{wmmmt2024} and for general gapped systems in \cite{mtw2025} (Theorem~\ref{thm:equiv}).
\end{itemize}

 As mentioned above, we prove the all of these statements also 
for a large class of generalized step functions (see Definition~\ref{def:genswitch}). For example, one can replace the   $\La_1$ and $\La_2$  by a pair of characteristic functions on  half-planes defined by any two transversal lines, including steps along lines with irrational slopes. More generally, one can increase the chemical potential on any cone-like region and consider the current into another cone-like region instead.

The structure of the proof of \eqref{main_intro} is similar to the one of \cite{wmmmt2024}: In a first step we use the defining properties of the NEASS $\w_\eps$ to show that $\Delta J_{\eps} =  \eps \, \w_\eps( \i\,[\, \La_2\ODeps,\, \La_1\ODeps\,]) + \mathcal{O}(\eps^\infty)$. Then we prove a Chern-Simons type lemma that shows that the quantity  $\w_\eps( \i\,[\, \La_2\ODeps,\, \La_1\ODeps\,])$ is invariant under locally generated automorphisms and thus, in particular,  independent of $\eps$.

\begin{SCfigure}[2][ht]
    \label{fig:currentdensity}
\usetikzlibrary{patterns}
\vspace{10mm}

\hspace{8mm}
 \begin{tikzpicture}[scale=.85]
  \begin{scope}
     \shade[shading=axis,                 
         left color=white, 
         right color=blue] 
    (-3.5,-3.5) rectangle (3.5,3.5);
  \end{scope}

  \draw[->,thick] (-3.7,0) -- (4,0) node[above] {\large$x_1$};
  \draw[->,thick] (0,-3.7) -- (0,4) node[right] {\large$x_2$};

    \draw[color=nred,->,ultra thick] (0,-.6) -- (0,.6);
    \draw[color=nred,->,ultra thick] (0,.9) -- (0,2.1);
    \draw[color=nred,->,ultra thick] (0,2.4) -- (0,3.6);
    \draw[color=nred,->,ultra thick] (0,-2.1) -- (0,-.9);
    \draw[color=nred,->,ultra thick] (0,-3.6) -- (0,-2.4);      
    
    \draw[color=nred,->,ultra thick] (1,-.6) -- (1,.6);
    \draw[color=nred,->,ultra thick] (1,.9) -- (1,2.1);
    \draw[color=nred,->,ultra thick] (1,2.4) -- (1,3.6);
    \draw[color=nred,->,ultra thick] (1,-2.1) -- (1,-.9);
    \draw[color=nred,->,ultra thick] (1,-3.6) -- (1,-2.4);      
    
        \draw[color=nred,->,ultra thick] (2,-.6) -- (2,.6);
    \draw[color=nred,->,ultra thick] (2,.9) -- (2,2.1);
    \draw[color=nred,->,ultra thick] (2,2.4) -- (2,3.6);
    \draw[color=nred,->,ultra thick] (2,-2.1) -- (2,-.9);
    \draw[color=nred,->,ultra thick] (2,-3.6) -- (2,-2.4);      
    
        \draw[color=nred,->,ultra thick] (3,-.6) -- (3,.6);
    \draw[color=nred,->,ultra thick] (3,.9) -- (3,2.1);
    \draw[color=nred,->,ultra thick] (3,2.4) -- (3,3.6);
    \draw[color=nred,->,ultra thick] (3,-2.1) -- (3,-.9);
    \draw[color=nred,->,ultra thick] (3,-3.6) -- (3,-2.4);      
    
      \draw[color=nred,->,ultra thick] (-1,-.6) -- (-1,.6);
    \draw[color=nred,->,ultra thick] (-1,.9) -- (-1,2.1);
    \draw[color=nred,->,ultra thick] (-1,2.4) -- (-1,3.6);
    \draw[color=nred,->,ultra thick] (-1,-2.1) -- (-1,-.9);
    \draw[color=nred,->,ultra thick] (-1,-3.6) -- (-1,-2.4);      
    
        \draw[color=nred,->,ultra thick] (-2,-.6) -- (-2,.6);
    \draw[color=nred,->,ultra thick] (-2,.9) -- (-2,2.1);
    \draw[color=nred,->,ultra thick] (-2,2.4) -- (-2,3.6);
    \draw[color=nred,->,ultra thick] (-2,-2.1) -- (-2,-.9);
    \draw[color=nred,->,ultra thick] (-2,-3.6) -- (-2,-2.4);      
    
        \draw[color=nred,->,ultra thick] (-3,-.6) -- (-3,.6);
    \draw[color=nred,->,ultra thick] (-3,.9) -- (-3,2.1);
    \draw[color=nred,->,ultra thick] (-3,2.4) -- (-3,3.6);
    \draw[color=nred,->,ultra thick] (-3,-2.1) -- (-3,-.9);
    \draw[color=nred,->,ultra thick] (-3,-3.6) -- (-3,-2.4);      

   \node at (2,-1) {\Large$\eps X_1$};
   
\end{tikzpicture}

\caption{
In contrast, the current response to a uniform electric field  of strength $\eps$ in the $x_1$-direction -- that is to a linear potential $\eps X_1$ -- is a uniform Hall current density. Adding such a perturbation closes the spectral gap of $H_0$  for any $\eps>0$. For periodic Hamiltonians it is shown in \cite{wmmmt2024}  and for general Hamiltonians in \cite{mtw2025}  that the current density is nearly linear in $\eps$, with the linear coefficient being the Hall conductivity, which is given by the expectation of a double commutator of modified position operators.
Since the current density in any gapped ground state is zero, there is no need to subtract a ground state contribution when defining the current density induced by an external field. 
}
\vspace{-10mm}
\end{SCfigure}
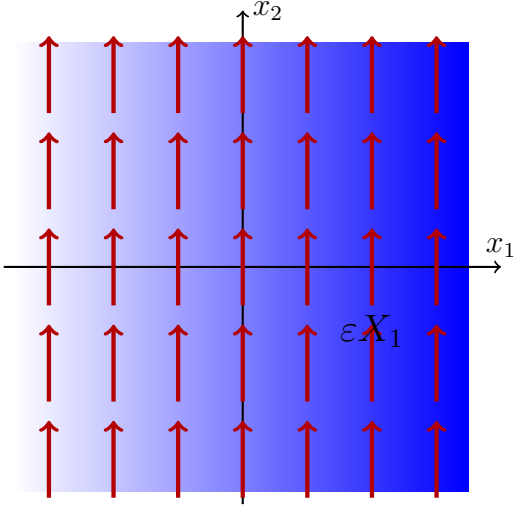

As previously mentioned, we expect that for sufficiently small values of $\eps$, the NEASS state $\w_\eps$ is still a gapped ground state of $H_\eps$, i.e.\ that an increase in the chemical potential in some region of space, if  small enough, does not close the spectral gap. For the non-interacting case, $H_0 = \mathrm{d} \Gamma(h_0)$, this is a straightforward consequence of perturbation theory at the level of the corresponding one-body Hamiltonian $h_0$. And for weakly interacting systems this follows from the stability of the spectral gap proved in \cite{de2019persistence,Ha19} and arguments along the line of the proof of Proposition~3.4 in \cite{wmmmt2024}. This shows that we cannot expect the current  into the upper half plane to vanish for generic gapped ground states $\w_0$, even in cases where $\w_0(\i[H_0, \Lambda_2]):= \sum_{ x \in \Z^2}  \w_0(\i[H_0,\Lambda_2]_x) $ is well defined. Thus, we need to consider the difference $\Delta J_\eps$,
which differs from the situation in a finite volume $[0,L]^2\cap\Z^2$ with periodic boundary conditions considered   in \cite{bachmann2021exactness}. In this case, Bloch's theorem states that the total current through any closed loop in a gapped ground state vanishes for $L\to \infty$, cf.\ \cite{BBDF20}.
However,  in Theorem~\ref{thm:nomacro} we show an infinite volume version of Bloch's theorem, namely that no macroscopic currents can flow in a gapped ground state:  For any gapped ground state $\w_0$ the current density across the line $\{x_2=0\}$ vanishes,
\[
    \lim_{k\to\infty}\frac{1}{2k+1}\sum_{x \in B_k(z)}  \w_0(\i[H_0,\Lambda_2]_x)\;=\;0  \quad \forall z\in \Z^2 \,,
\]
where for $k\in\N$ we define $B_k(z) \coloneq \{ x \in \Z^2 \, | \, \lVert x\-z \rVert \leq k \} $ as the box of side length $2k$ around $z$.  Also here, the same is true for currents across the boundaries of a class of cone-like regions.
This contrasts with the situation in which a uniform electric field is applied in the $x_1$-direction, closing the spectral gap and leading to a NEASS state $\w_\eps$ which carries a non-vanishing current density in the $x_2$-direction, see Figure \ref{fig:currentdensity}.

Finally, let us  compare the NEASS approach to  to the charge pump approach used in \cite{bachmann2021exactness} in some more detail. Both approaches start from a definition of the Hall conductance $\sigma_\mathrm{H}$ based  on adiabatic switching. More precisely, consider the time-dependent Hamiltonian
\[
H_{\eps,\eta} (t) :=  H_0 + \eps f(\eta t) \Lambda_1
\]
for a smooth function $f:\R \to \R$ with supp$f'\subset [0,1]$ and $f(0)=0$. Here $0<\eta\ll 1$ is the adiabatic parameter.
Let $\mathfrak{U}_{0,t}^{\eps,\eta}$ denote the evolution family generated by the time-dependent Hamiltonian $H_{\eps,\eta}(t)$ with starting time $t_0=0$, which in our setting is a cocycle of   automorphisms of the quasi-local algebra.

In the NEASS approach one assumes  in addition that $f_\mathrm{NE }(1)=1$, i.e.\ one adiabatically turns on the perturbation during the time interval $[0,\eta^{-1}]$, and then considers, as explained above, the additional current flowing in the state $\w_0\circ \mathfrak{U}^{\eps,\eta}_{0,t}$  that is reached be the adiabatic evolution  for times $t\geq\eta^{-1}$ when the perturbation is fully switched on. It is shown in \cite{Teufel2020,HenheikTeufel2022,BTW2025} that for $\eta = \eps^\alpha$ with some $\alpha>0$ and all $t\geq\eta^{-1}$ the state $\w_0\circ \mathfrak{U}^{\eps,\eta}_{0,t}$ is $\mathcal{O}(\epsi^\infty)$-close to the NEASS state $\w_\eps$ considered in this paper. Now recall that in general a conductance is defined as the ratio of current to voltage drop in the limit of a small voltage drop. In the NEASS approach the current is $\Delta J_\eps$ and the voltage drop is $\eps$, so  
\[
\sigma_\mathrm{H}^\mathrm{NE} := \lim_{\eps\to0} \frac{\Delta J_\eps}{\eps}\,.
\]
Our main result \eqref{main_intro} thus shows that $\sigma_\mathrm{H}^\mathrm{NE} =  \w_0( \i\,[\, \La_2\OD,\, \La_1\OD\,])$ and that
\[
\left|\frac{\Delta J_\eps}{\eps} - \sigma_\mathrm{H}^\mathrm{NE}\right|=\mathcal{O}(\eps^\infty)\,,
\]
i.e.\ that for small $\eps$ 
the finite voltage ratio $\frac{\Delta J_\eps}{\eps}$ is a very good approximation to $\sigma_\mathrm{H}^\mathrm{NE}$.

In the charge pump approach (see \cite{klein1990power,bachmann2021exactness}), the perturbation is turned on and off again adiabatically using a switching function
$f_\mathrm{CP}$ with $f_\mathrm{CP}(0)=0=f_\mathrm{CP}(1)$ and $\int_0^1f(t)\,\mathrm{d} t = 1$. One then considers the total charge transported into the upper half-plane during one cycle,
\[
\Delta Q_{\eps,\eta}:= (\w_0\circ \mathfrak{U}^{\eps,\eta}_{0,\eta^{-1}} - \w_0)(\Lambda_2)\,.
\]
In order to relate $\Delta Q_{\eps,\eta}$ to a conductance, one considers
  the instantaneous current flowing into the upper half plane at time $t$ defined by the time derivative of the total charge in the upper half plane, i.e.\ by $\frac{\mathrm{d}}{\mathrm{d}t} \,\w_0\circ \mathfrak{U}^{\eps,\eta}_{0,t}(\Lambda_2)$. Now set  $\eps=\eta$ and define $\sigma_\mathrm{H}^\mathrm{CP}$ as the ratio of the time-averaged current and the time-averaged voltage,
\[
\sigma_\mathrm{H}^\mathrm{CP} := \lim_{\eps\to 0} \frac{\eps \int_0^{\eps^{-1}}\frac{\mathrm{d}}{\mathrm{d}t} \,\w_0\circ \mathfrak{U}^{\eps,\eps}_{0,t}(\Lambda_2)\,\mathrm{d}t }{\eps\int_0^{\eps^{-1}} \eps f(\eps t) \,\mathrm{d}t} = \lim_{\eps\to 0} \Delta Q_{\eps,\eps} \, .
\]
In \cite{bachmann2021exactness} and \cite{BBDF20} it is shown, for the appropriate finite volume version of this setup on a torus, that actually
\[
\Delta Q_{\eps,\eps}= \w_0( \i\,[\, \La_2\OD,\, \La_1\OD\,])+ \mathcal{O}(\epsi^\infty) 
\]
and thus
\[
\sigma_\mathrm{H}^\mathrm{CP} = \sigma_\mathrm{H}^\mathrm{NE}=\w_0( \i\,[\, \La_2\OD,\, \La_1\OD\,])\quad\mbox{and}\quad
\left|\Delta Q_{\eps,\eps} - \sigma_\mathrm{H}\right| = \mathcal{O}(\eps^\infty)\,.
\]
The proof in \cite{bachmann2021exactness} is based on the many-body version of the adiabatic theorem (see \cite{bachmann2018adiabatic} and also \cite{monaco2019adiabatic}) and assumes that  $H_{\eps,\eta}(t)$ has a gapped ground state for all $t\in[0,\eta^{-1}]$. Although we previously argued that we expect that the gap will remain open    when   $\eps \Lambda_1$ is added to a gapped $H_0$ for $\eps$ small enough, this has not yet been proven for general gapped $H_0$.
In summary, although the two approaches start from different definitions of Hall conductance, they agree on its value. However, the NEASS approach shows that additional time averaging is unnecessary and does not rely on the assumption that the gap does not close.

\section{Mathematical setup}\label{sec:setup}

The anti-symmetric  (or fermionic) Fock space over the lattice $\Z^2$ is given by
    \begin{align*}
       \mathcal{F}(\Z^2,\C^n) := \bigoplus_{N=0}^{\infty}\ell^2(\Z^2,\C^n)^{\wedge N} ~.
    \end{align*}
    We use $a^*_{x,i}$ and $a_{x,i}$ for $x\in \Z^2$, $i\in \{1,\dots,n\}$ to denote the fermionic creation and annihilation operators associated to the standard basis of $\ell^2(\Z^2,\C^n)$ and recall that they satisfy the canonical anti-commutation relations (CAR).
    The number operator at site $x\in\Z^2$ is defined by
    \begin{align*}
        n_x:=\sum_{i=1}^n a^*_{x,i}a_{x,i}\,.
    \end{align*}
    The algebra of all bounded operators on $\mathcal{F}(\Z^2,\C^n)$ is denoted by $\mathcal{B}(\mathcal{F}(\Z^2,\C^n))$. 
    For each $M\subseteq \Z^2$ let $\mA_M$ be the C*-subalgebra of $\mathcal{B}(\mathcal{F}(\Z^2,\C^n))$ generated by
    \begin{align*}
        \{a^*_{x,i}~|~x\in M,~ i\in \{1,\dots,n\}\}~.
    \end{align*}
    The C*-algebra $\mA := \mA_{\Z^2}$ is the CAR-algebra, which we also call the quasi-local algebra. We write $P_0(\Z^2) := \{M\subseteq \Z^2~|~|M|<\infty \}$ and call
    \begin{align*}
        \mA_0 := \bigcup_{M\in P_0(\Z^2)} \mA_M \subseteq \mA
    \end{align*}
    the local algebra. Consequently, an operator is called quasi-local if it lies in $\mA$ and local if it lies in $\mA_0$.
    For each $\varphi \in \R$ there is a unique automorphism\footnote{In the following the term {\it automorphism} is used in the sense of a $*$-automorphism as defined in the textbook \cite{bratteliI}.} $g_\varphi$ of $\mA$, such that
    \begin{align*}
        g_\varphi(a^*_{x,i}) =  \e^{i\varphi} \, a^*_{x,i}, \quad \text{for all}~~ x\in \Z^2,~ i\in \{ 1,\dots,n \} ~.
    \end{align*}
    One defines the set
    \begin{align*}
        \mA^N := \{A\in \mA~|~  \forall \varphi \in \R :\, g_\varphi (A) = A \}
    \end{align*}
    and calls $\mA^N$ the gauge-invariant sub-algebra of $\mA$. It is the closure of the set of all local observables that commute with the number operator.  Its part in $M\subseteq\Z^2$  is denoted by $\mA^N_M:= \mA^N\cap \mA_M$.
    For disjoint regions $M_1,M_2 \subseteq\Z^2$,  $M_1\cap M_2 = \emptyset$, operators  $A\in \mA_{M_1}^N$ and $B\in \mA_{M_2}$ commute;  $[A,B] = 0$. Positive linear functionals $\w \colon \mA \to \C$ of norm $1$ are called states.

In order to define quantitative notions of localization for quasi-local operators, one makes use of the fact that one can localize operators to given regions by means of the fermionic conditional expectation. 
To this end first note that  $\mA$ has a unique state $\w^{\tr}$ that satisfies
\begin{align*}
    \w^{\tr}(AB) = \w^{\tr}(BA)
\end{align*}
for all  $A,B \in \mA$, called the tracial state (e.g.\ \cite[Definition 4.1, Remark 2]{ArakiMoriya2002}).

\medskip

\begin{proposition}[{\cite[Theorem 4.7]{ArakiMoriya2002},\cite[Proposition 2.1]{wmmmt2024}}]\label{Ex+UniqueExpectation}
    For each $M\subseteq\Z^2$ there exists a unique linear map 
    \begin{align*}
        \E_M:\mA \to \mA_M\,,
    \end{align*}
    called the conditional expectation with respect to $\w^\tr$, such that
    \begin{equation}\label{eq:Conditional expectation defining property}
      \forall A\in \mA \; \;\forall B\in \mA_M \,:\quad \w^{\tr}(AB)=\w^{\tr}(\E_M(A)B) \,.
    \end{equation}
    It is unital, positive and has the properties 
    \begin{eqnarray*}
        \forall M\subseteq \Z^2\;\; \forall A,C\in \mA_M\;\; \forall B\in\mA\, : &&   \E_M \br{A\,B\,C} = A\, \E_M(B)\,C\\[1mm]
      \forall M_1,M_2 \subseteq \Z^2\,:&&   \E_{M_1} \circ \E_{M_2}  = \E_{M_1\cap M_2}\,\\[1mm]
        \forall M \subseteq \Z^2\,:&& \E_M \mA^N \subseteq \mA^N
  \\[1mm]
        \forall M \subseteq \Z^2 \; \;  \forall A \subseteq \mA\,:&&  \norm{\E_M(A)}\leq \norm{A}\,.
    \end{eqnarray*}
\end{proposition}

\medskip

With the help of $\E$ we can define subspaces of $\mA^N$ that contain operators with well-defined decay properties (cf.\ \cite{MO20} for similar definitions for quantum spin systems).

\medskip

\begin{definition}
    For $\nu \in \N_0$, $x \in \Z^2$ and $A \in \mA$ let
    \begin{align*}
        \norm{A}_{\nu,x} \coloneq \norm{A} + \sup_{k\in \N_0} \norm{A-\E_{B_k(x)}A} \, (1+k)^\nu \, ,
    \end{align*}
    where $B_k(x) \coloneq \{y\in \Z^2 \, | \, \norm{x-y} \leq k \}$ is the box with side-length $2k$ around $x$ with respect the maximum norm on $\Z^2$.
    We denote the set of all gauge-invariant $A\in \mA^N$ with finite $\norm{\cdot}_{\nu,x}$ for some (and therefore all) $x\in \Z^2$ and all $\nu\in \N_0$ by $D_\infty$.
\end{definition}

\medskip

\begin{lemma}\label{lem:commutator bound}
    Let $A,B \in D_\infty$, $\nu,m \in \N_0$ and $x,y \in \Z^2$. It holds that
    \begin{align*}
        \norm{[A,B]}_{\nu,x} \leq 4^{\nu+m+3} \frac{\norm{A}_{\nu+m,y}\, \norm{B}_{\nu+m,x}}{(1+\norm{x-y})^m}\, .
    \end{align*}
    Here the norm on $\Z^2$ is again the maximum norm.
\end{lemma}
\begin{proof}
    Let $\Z^2 \to \aut(\mA), \, \g \mapsto T_\g$ be the standard translation on $\mA$. We observe that 
    \begin{align*}
        \norm{[A,B]}_{\nu,x} &= \norm{T_{-x}[A,B]}_{\nu,0} = \norm{[T_{y-x}T_{-y}A,T_{-x}B]}_{\nu,0}\, .
    \end{align*}
    Now using the bound from Lemma A.1 in \cite{wmmmt2024} we get
    \begin{align*}
        \norm{[A,B]}_{\nu,x} &\leq 4^{\nu+ m + 3} \frac{\norm{T_{-y}A}_{\nu+m,0}\, \norm{T_{-x}B}_{\nu+m,0}}{(1+ \norm{x-y})^m}\\
        &=4^{\nu+ m + 3} \frac{\norm{A}_{\nu+m,y}\, \norm{B}_{\nu+m,x}}{(1+ \norm{x-y})^m} \, . \qedhere
    \end{align*}
\end{proof}

\medskip

The relevant physical dynamics on $\mA$ is generated by densely defined derivations, which in turn are constructed from so-called interactions.
    An interaction is a map $\Phi: P_0(\Z^2) \to \mA^N$, such that $\Phi(\emptyset) = 0$ and for all $M\in P_0(\Z^2)$ it holds that $\Phi(M) \in \mA_M$, $\Phi(M)^* = \Phi(M)$, and the sum 
    \begin{equation}\label{eq:uncond}\sum_{\substack{K\in P_0(\Z^2)\\ M\cap K \neq \emptyset}} \Phi(K)
    \end{equation}
    converges unconditionally.  Note that in our definition the local terms of an interaction are always gauge-invariant.
For two interactions   $\Phi$ and $\Psi$ their commutator is given by
    \begin{align*}
        [\Phi,\Psi]: P_0(\Z^2) \to \mA^N\,,\qquad M\mapsto 
        [\Phi,\Psi](M) \;:= \sum_{\substack{M_1,M_2 \subseteq M\\ M_1 \cup M_2= M}} [\Phi(M_1),\Psi(M_2)]
    \end{align*}
    and the map $\i[\Phi,\Psi]$ satisfies the definition of an interaction except for unconditional convergence as in~\eqref{eq:uncond}, which is not always satisfied. All the commutators of interactions appearing in the following will however be interactions. 
Interactions  define derivations on the algebra in the following way.    
    For an interaction $\Phi$ let
    \begin{align*}
        \mL_{\Phi}^\circ: \mA_0 \to \mA,~ A\mapsto \sum_{M\in P_0(\Z^2)} [\Phi(M),A] \, .
    \end{align*}
    It follows from \cite[Proposition 3.2.22]{bratteliI} and \cite[Proposition 3.1.15]{bratteliI} that $\mL_{\Phi}^\circ$ is closable. We denote its closure by $\mL_{\Phi}$ and call it the Liouvillian of $\Phi$.
Some basic interactions that will appear frequently in the following are the number operator $N$, the position operators $X_j$ for $j\in\{1, 2\}$, and the half-plane step functions $\Lambda_j$ for $j\in\{1, 2\}$, which are non-vanishing only on one-element sets. They are defined by $N(\{x\}) = n_x$, $X_j(\{x\}) = x_j \, n_x$, and 
$$
\Lambda_j(\{x\})= \begin{cases}
    n_x & x_j \geq 0\\
    0 & \text{else}
\end{cases}
$$  for $x\in\Z^2$. The interactions $\Lambda_1$ and $\Lambda_2$ are instances of a class of interactions defined by \emph{generalized step functions}, which are bounded functions $f\colon \R^2 \mapsto \R$ together with a choice of complementary disjoint open cones $C_1$ and $C_0$, such that $f|_{C_0} = 0$ and $f|_{C_1} = 1$ (see Definition~\ref{def:genswitch} and Remark \ref{rem: half-plane step functions}).

\medskip

\begin{definition}\label{norm interaction}
    Let $\Phi$ be an interaction and $\nu \in \N$. Let
    \begin{align*}
        \|\Phi\|_{\nu} := \sup_{x\in \Z^2} \sum_{\substack{M\in
        P_0(\Z^2)\\x\in M}}(1+\mathrm{diam}(M))^\nu  \| \Phi(M)\|  \, ,
    \end{align*}
    where $\mathrm{diam}(M)$ is the maximal distance of two elements in $M$ with respect to the maximum norm on $\Z^2$.
    The set of all interactions with finite $\norm{\cdot}_{\nu}$ for all $\nu \in \N_0$ is denoted by $B_{\infty}$. \\
    For an interval $I\subseteq \R$, we define $B_{\infty,I}^{(0)}$ as the set of families of $B_\infty$-interactions $(\Phi^s)_{s \in I}$ where for each $M\in P_0(\Z^2)$, the map $ s \mapsto \Phi^s(M) $ is continuous that and that satisfy $\sup_{s \in I}\norm{\Phi^s}_{\nu} < \infty$ for all $\nu \in \N_0$.
\end{definition}

\medskip

\begin{definition}\label{def: quasi-local terms}
    For $M \in P_0(\Z^2)$ we define its center $\mathrm{C}(M) \in M$ as the point in $M$ that minimizes the distance to its center of mass $\mathrm{cm}(M)\in\R^2$. If there are multiple such points we choose the one where the angle $$\angle(e_1,\mathrm{C}(M)-\mathrm{cm}(M))\in [0,2\pi)$$ is minimal, with $e_1$ being the unit vector in the first coordinate direction.
    Let $x\in \Z^2$. We define $R_x \subseteq P_0(\Z^2)$ to be set of all finite subsets of $\Z^2$ that have their center in $x$. Given an interaction $\Phi$ we define 
    \begin{align*}
        \Phi_x \coloneq \sum_{M \in R_x} \Phi(M) \, .
    \end{align*}
\end{definition}

\medskip

\begin{remark}
    Note that for any interaction $\Phi$ and lattice point $x$ the quasi-local observable $\Phi_x$ is well-defined and that $ \lVert \Phi_x \rVert_{\nu,x} \leq 3 \, \lVert \Phi \rVert_\nu $ for all $\nu \in \N_0$.
\end{remark}

\medskip

\begin{lemma}\label{lem: sum representation of generator}
    For an interaction of the form $\Psi = p\, \Phi + q \, X_j$, where $p,q\in\R$, $j\in \{1,2\}$ and $\Phi \in  B_\infty$, it holds that $D_\infty \subseteq D(\mL_{\Psi})$. Let $ A\in D_\infty$, then the sums 
    \[\sum_{M\in P_0(\Z^2)} [\,\Psi(M) , \,A\,] \quad \text{and} \quad \sum_{x \in \Z^2} [\, \Psi_x , \, A\,] \] converge absolutely
    and
    \[ 
    \mL_{\Psi }A 
    \;=\;  \sum_{M\in P_0(\Z^2)} [\,\Psi(M),\, A\,] 
    \;=\;  \sum_{x\in \Z^2} [\, \Psi_x, \, A \, ]
    \,. 
    \]
    For each $\nu \in \N_0$ there is a constant $c_\nu$, independent of $\Phi$ and $ A $, such that for all~$x\in \Z^2$
    \[ \norm{ \mL_{\Psi } \, A }_{\nu,x} \leq c_\nu \, (p \,\norm{\Phi}_{3+2\nu} + q \,\norm{n_0}) \, \norm{A}_{5+2\nu}\,.\]
\end{lemma}
\begin{proof}
    The statement follows from Lemma 2.5 in \cite{wmmmt2024} in the following way: The convergence of the first sum is directly implied by the lemma and the second sum results from choosing a specific summation order. To get the estimate, denote by $T$ the family of standard lattice-translation automorphisms acting on $\mA$. It holds that
    \begin{align*}
        \norm{ \mL_{\Psi } \, A }_{\nu,x} = \norm{ T_{-x} \, \mL_{\Psi } \, A }_{\nu,0} = \norm{  \mL_{T_{-x}\Psi } \, T_{-x}\, A }_{\nu,0}\, .
    \end{align*}
    Lemma 2.5 of \cite{wmmmt2024}, together with the fact that, due to gauge-invariance of $A$, we have $\mL_{T_{-x}X_j}\,A = \mL_{X_j}\,A$, tells us that there is a constant $c_\nu$ independent of $x$, $\Phi$ and $A$ such that
    \begin{equation*}
        \norm{  \mL_{T_{-x}\Psi } \, T_{-x}\, A }_{\nu,0} \leq c_\nu \, (p \,\norm{T_{-x}\, \Phi}_{3+2\nu} + q \,\norm{n_0}) \, \norm{T_{-x}\,A}_{5+2\nu} = c_\nu \, (p \,\norm{\Phi}_{3+2\nu} + q \,\norm{n_0}) \, \norm{A}_{5+2\nu} \,. \qedhere
    \end{equation*}
\end{proof}

\medskip

\begin{definition}
    Let $I\subseteq \R$ be an interval. A family $(\alpha_{u,v})_{(u,v) \in I^2}$ of automorphisms of $\mA$ is called a \emph{cocycle} if it satisfies
    \begin{align*}
        \forall t,u,v \in I, \quad \alpha_{t,u} \, \alpha_{u,v} =\alpha_{t,v}\, .
    \end{align*}
    We say the cocycle is generated by the family of interactions $(\Phi^{v})_{v\in I}$, if for all $A \in \mA_0$ it holds that 
    \begin{align*}
        \partial_v \, \alpha_{u,v} \, A = \alpha_{u,v} \, \i \, \mL_{\Phi^{v}} \, A  \, .   
    \end{align*}
    We call $(\alpha_{u,v})_{(u,v) \in I^2}$ locally generated if it is generated by some family of interactions in $B_{\infty,I}^{(0)}$. An automorphism is called locally generated if it is part of a locally generated cocycle.
\end{definition}

\medskip

The following Lemma states that locally generated automorphisms worsen the decay of a quasi-local observable only by a polynomially growing constant. It is based on Lieb-Robinson bounds shown in \cite{TeufelWessel25}.

\medskip
\begin{lemma}[\cite{BTWautomorphic} Lemma 2.10]\label{lem: cocycles}
    Let $I \subseteq \R$ be an interval and $(\Phi^{v})_{v\in I}$ a family of interactions in $B_{\infty,I}^{(0)}$. Then there exists a unique cocycle of automorphisms $(\alpha_{u,v})_{(u,v) \in I^2}$ generated by $(\Phi^{v})_{v\in I}$.
    And for all $\nu \in \N_0$ there exists an increasing function $b_\nu:\R \to \R$, that grows at most polynomially, such that
    \begin{align*}
        \norm{\alpha_{u,v} \, A}_{\nu,x} \leq b_\nu(|v-u|) \norm{A}_{\nu,x} \,   \quad \text{ for all } x\in \Z^2, \, A\in \mA_\infty,\, u,v\in I \,.
    \end{align*}
    Furthermore, for all  constants $C>0$ the function $b_\nu$ can be chosen uniformly for all $(\Phi^v)_{v\in I}$ \linebreak with $\sup_{s\in I} \lVert \Phi^v \rVert_{4\nu+22} < C$.
\end{lemma}

\section{Gapped response systems}

In the following, a (topological) insulator is described by an interaction $H\in B_{\infty}$ that has at least one gapped ground state $\omega_0$.  

\medskip

\begin{definition}\label{def:groundstate}
Let    $H\in B_{\infty}$. We say that a state $\omega_0$ is a  \emph{gapped ground state} for $H$ with gap  $g>0$, iff   for all $A\in \mA_0$  
    \begin{equation}\label{eq:gap}
    \w_0( A^* \mL_H A ) \geq g \left( \w_0(A^* A) - \left| \w_0(A) \right|^2 \right)\,.
    \end{equation}
This condition implies that in the GNS representation induced by $\w_0$ the GNS Hamiltonian representing $\mL_H$ has a unique gapped ground state. For this reason states satisfying \eqref{eq:gap} are also called locally unique gapped ground states.
\end{definition}

 We are interested in how a system that is initially in a gapped ground state responds to a small change in the chemical potential in the right half-plane, or more generally a cone-like region.
Assuming that the system is initially in  a ground state $\omega_0$, 
it is shown in \cite{Teufel2020,HenheikTeufel2022,BTW2025}
that when adiabatically switching on  the perturbation $\eps \Lambda_f$, where $f$ is a generalized step function (see Definition \ref{def:genswitch} ),  the system evolves (up to errors of order  $\mathcal{O}(\eps^{\infty})$) into a state $\w_\eps := \w_0 \circ \beta_\eps$ (called NEASS), where $\beta\colon \R \to \mathrm{Aut}(\mA),~ \eps \mapsto \beta_{\eps}$, is a continuous one-parameter family of locally generated automorphisms.

Our main result concerns the change in the current flowing into the upper half-plane, or more generally another cone-like region, as the system evolves from $\w_0$ to $\w_\eps$.
Since our analysis does not depend on the way   a NEASS is constructed or reached by adiabatic driving, we formulate the relevant properties of NEASSs  in a definition  and refer exclusively to these properties in all main results. The existence of a NEASS with these properties, and the fact that such a NEASS is reached by adiabatic driving, is proved in \cite{BTW2025}.

\medskip

\begin{definition}\label{def:NEASS}
   Let $H\in B_{\infty}$, let   $\omega_0$ be a gapped ground state for $H$, and $f$ a generalized step function. We say that a family $[-1,1]\ni\eps \mapsto \w_\eps = \w_0 \circ  \e^{\i \mL_{S^{\eps}}}$ is a \emph{NEASS} for $H$, $\omega_0$, and $\Lambda_f$, iff  $[-1,1]\ni \eps \mapsto S^{\eps}$ is a family of   interactions $S^{\eps}\in B_\infty$ such that 
   for all $\nu,m\in\N_0$
   \begin{equation}\label{eq:Sloc}
         \sup_{\eps \in [-1,1]}  \lVert S^\eps \rVert_{\nu} < \infty \, ,\qquad \sup_{\eps \in [-1,1]} \sup_{x\in \Z^2} \lVert S^\eps_x \rVert_{\nu,x} \, \left(1 + \mathrm{d}(x,C^f_\delta)\right)^m < \infty \, ,
   \end{equation}
   where $C^f_\delta = \R^2\setminus (C_0^f\cup C_1^f)$, and
   \begin{equation}\label{NEASSinvariant}
    \sup_{x\in\Z^2}\sup_{A\in D_\infty\setminus \{0\}} \frac{\lvert \w_\eps(\mL_{H+\eps\La_f}\,A)\rvert }{\lVert A \rVert_{6,x}} = \mathcal{O}(\eps^\infty) \, .
    \end{equation}
\end{definition}

The second equation in \eqref{eq:Sloc} states that the generator $S^\eps$ of the NEASS transformation is localized near the region $C^f_\delta$ where the step function $f$ is not necessarily constant.
Equation~\eqref{NEASSinvariant} states that $\omega_\eps\circ \mathcal{L}_{H_\eps}$ is small in a sufficiently strong sense, and hence  that $\omega_\eps$ is almost invariant for the dynamics generated by~$H_\eps:=H+\eps \Lambda_f$ for long times.

In  Section~\ref{sec:off-diagonal} we define the off-diagonal part of observables and interactions, which involves the choice of a weight function $W$ with certain properties. This choice   affects some objects in the statement of Theorem~\ref{thm:main}. The resulting Hall conductance is, however, independent of this choice, which is also part of the theorem. We make all choices explicit and summarize them together with our assumptions in the following definitions:

\medskip

\begin{definition}
    Let $H$ be an interaction in $B_{\infty}$, that has a gapped ground state $\w_0$ with gap $g>0$. Further let $W \colon \R \to \R$ be a function that satisfies:
    \begin{itemize}
        \item[(i)] $W$ is odd.
        \item[(ii)] the Fourier transform $\widehat W$ of $W$ satisfies $\widehat{W}(k) = \frac{- \i}{\sqrt{2  \pi}\, k}$ for  $k \in \R \setminus [-g,g]$.
        \item[(iii)] $\sup_{s\in\R} \lvert s \rvert^n \, W(s) < \infty$ for all $n\in \N_0$.
    \end{itemize}
    We then call the tuple $(\w_0,\,H,\, W)$ a \emph{gapped system}. Note that such a function $W$   exists for any $g>0$ (see \cite[Lemma 2.6]{BMNS12}). 
    
    Additionally, let $f$ be a generalized step function and $(S^\eps)_{\eps\in [-1,1]}$ a family of interactions that defines a NEASS for $H$, $\w_0$ and $\Lambda_f$ in the sense of Definition~\ref{def:NEASS}.
    Then we call the tuple $(\w_0,\,H,\, W,\, (S^\eps)_{\eps\in [-1,1]})$ a \emph{gapped response system with respect to $f$}. 
    
    Given a gapped (response) system, we always use the following notations: For any automorphism $\alpha$ we write $\w_\alpha \coloneq \w_0 \circ \alpha$, we define $\beta_\eps \coloneq \e^{\i \mL_{S^\eps}}$, and use $\w_\eps \coloneq  \w_0 \circ \beta_\eps$ to denote the NEASS family. And we denote the quasi-local terms of the Hamiltonian by $h_x$ instead of $H_x$ for $x\in \Z^2$ (see Definition \ref{def: quasi-local terms}).  
\end{definition}

\section{The off-diagonal map}\label{sec:off-diagonal}

In this section we recall and slightly generalize a standard construction that associates with an operator resp.\ with an interaction its part that is off-diagonal part with respect to any gapped ground state  of a Hamiltonian $H$ with gap at least $g>0$ (see, e.g., \cite{MO20,wmmmt2024} and references therein).

\medskip

\begin{definition}
    Let $(\w_0,\,H,\, W)$ be a gapped  system. For a locally generated automorphism $\alpha$ and $A\in D_\infty$, we define
    \begin{align*}
        A\OD[\alpha] :=  - \i \,  \alpha^{-1} \int_{\R}\mathrm{d}s ~W(s)\, \e^{\i s\mL_{H}} \, \mL_{H} \, \alpha \, A~.
    \end{align*}
    If $\alpha = \mathrm{id}$, we write $A\OD[\alpha] = A\OD$. Given a gapped response system $(\w_0,\,H,\, W,\, (S^\eps)_{\eps\in [-1,1]})$ and $\alpha = \beta_\eps$, we write $A\OD[\alpha] = A\ODeps$.
\end{definition}

\medskip

\begin{lemma}\label{OD-property Observable}
    Let $(\w_0,\,H,\, W)$ be a gapped  system, $\alpha$ a locally generated automorphism  and $A, B \in D_\infty$. The off-diagonal map satisfies the off-diagonal property:
    \begin{equation*}
        \w_\alpha([\,A , \, B \, ] ) = \w_\alpha([\,A\OD[\alpha] , \, B \, ] ) \, .
    \end{equation*}
\end{lemma}
\begin{proof}
    The statement is a straight forward consequence of Proposition 3.3 of \cite{HenheikTeufel2022}.
\end{proof}

\medskip
 
\begin{remark}
    The  notion  ``off-diagonal'' is motivated by the fact that in the setting of  Lemma \ref{OD-property Observable} one can  define 
    \[A\DI[\alpha]:= A-A\OD[\alpha]\]
    and finds that
    \[
        \w_\alpha(A) = \w_\alpha(A\DI[\alpha]) \quad\mbox{and}\quad \w_\alpha([\,A\DI[\alpha] , \, B \, ] )=0 \,.
    \] 
\end{remark}

\medskip

\begin{definition}\label{def:OD}
    Given a gapped system $(\w_0,\,H,\, W)$, a locally generated automorphism $\alpha$, and an interaction of the form $\Psi= p\, \Phi + q\, X_j$, where $p,q\in \R$, $j\in \{1,2\}$, $\Phi \in B_\infty$. We define the quasi-local operator 
    \begin{align*}
        (\Psi\OD[\alpha])_{x,*} \coloneq \i \, \alpha^{-1} \int_{\R}\dd s \, W(s)\, \e^{\i s \mL_H} \, \alpha \, \mL_{\Psi} \, \alpha^{-1} \, h_x \, ,
    \end{align*}
    for each $x\in \Z^2$ and the interaction $\Psi\OD[\alpha]$ by
    \begin{align*}
        \Psi\OD[\alpha](B_k(x)) &= (\E_{B_k(x)} - \E_{B_{k-1}(x)})\, (\Psi\OD[\alpha])_{x,*}\\
        \Psi\OD[\alpha](B_0(x)) &= \E_{B_0(x)}\, (\Psi\OD)_{x,*}
    \end{align*}
    and $\Psi\OD[\alpha](M) = 0 $ if $M$ is not equal to $B_k(x)$ for any $(k,x)\in  \N_0\times  \Z^2$.

    If $\alpha = \mathrm{id}$, we write $\Psi\OD[\alpha] = \Psi\OD$. Given a gapped response system $(\w_0,\,H,\, W,\, (S^\eps)_{\eps\in [-1,1]})$ and $\alpha = \beta_\eps$, we write $\Psi\OD[\alpha] = \Psi\ODeps$.
\end{definition}

\medskip

\begin{lemma}\label{OD-property Interaction}
    Let $(\w_0,\,H,\, W)$ be a gapped system, $\alpha$ a locally generated automorphism, and let \linebreak $\Psi= p\, \Phi + q\, X_j$, where $p,q\in \R$, $j\in \{1,2\}$, $\Phi \in B_\infty$. It holds that:
    \begin{itemize}
        \item[\rm (i)] $\Psi\OD[\alpha]$ is a $B_\infty$-interaction.
        \item[\rm (ii)] $\forall \, x\in \Z^2$, $(\Psi\OD[\alpha])_x = (\Psi\OD[\alpha])_{x,*} $
        \item[\rm (iii)] $\forall \, A \in D_\infty $, $\mL_{\Psi\OD[\alpha]}\, A = \sum_{x\in \Z^2} [\, (\Psi_x)\OD[\alpha], \, A \, ]$
        \item[\rm (iv)] $\forall \, A \in D_\infty $, $\w_\alpha(\mL_{\Psi}\, A) = \w_\alpha(\mL_{\Psi \OD[\alpha]}\, A)$
    \end{itemize}
\end{lemma}
\begin{proof}
    We first establish $\sup_{x\in \Z^2} \lVert (\Psi\OD[\alpha])_{x,*} \rVert_{\nu,x} <\infty$ for all $\nu \in \N_0$. This follows from Lemma \ref{lem: cocycles} and Lemma \ref{lem: sum representation of generator} together with the fact that $\sup_{x\in \Z^2} \lVert h_x \rVert_{\nu,x} \leq 3\lVert H \rVert_\nu <\infty$. Form here it is easy to see that $\Psi\OD[\alpha]$ lies in $B_\infty$:
    \begin{align*}
        \hspace{2em}&\hspace{-2em} \sum_{\substack{M\in P_0(\Z^2)\\x\in M}} (1+\diam(M))^\nu \, \lVert \Psi(M) \rVert \;=\\
        & = \;\sum_{k=1}^\infty \sum_{\substack{y\in \Z^2 \\ \lVert x-y \rVert = k }} (1+ 2\, k)^\nu \, \lVert (\E_{B_k(y)}-\E_{B_{k-1}(y)}) \, (\Psi\OD[\alpha])_{y,*} \rVert + \lVert \E_{B_0(x)} \, (\Psi\OD[\alpha])_{x,*} \rVert\\
        &\leq \;\sum_{k=1}^\infty \sum_{\substack{y\in \Z^2 \\ \lVert x-y \rVert = k }} \frac{2\,(1+ 2\, k)^\nu}{k^{\nu+3}} \, \lVert (\Psi\OD[\alpha])_{y,*} \rVert_{\nu+3,y} + \lVert (\Psi\OD[\alpha])_{x,*} \rVert\\
        &\leq \;\sum_{k=1}^\infty  \frac{2\,(1+ 2\, k)^\nu \, 8\, k}{k^{\nu+3}} \, \sup_{z \in \Z^2 } \lVert (\Psi\OD[\alpha])_{z,*} \rVert_{\nu+3,z} + \sup_{z \in \Z^2 } \lVert (\Psi\OD[\alpha])_{z,*} \rVert\, .
    \end{align*}
    Statement (ii) follows immediately from the fact that the center of a box $B_k(x)$ in the sense of Definition \ref{def: quasi-local terms} is $x$.
    To prove (iii), we first need to show that the the expression
    \begin{align*}
        \mL_{\Psi\OD[\alpha]} \, A = \sum_{x\in \Z^2} \int_{\R}\dd s \sum_{y\in \Z^2} \i[\, \alpha ^{-1} \,  \e^{\i s \mL_H} \, [\, \alpha \, \Psi_y, \, h_x \,],\, A\, ]
    \end{align*}
    (where we used Lemma \ref{lem: sum representation of generator}) converges absolutely. This can be accomplished with the bounds from Lemma \ref{lem: cocycles} and Lemma \ref{lem:commutator bound}. One can then exchange the order of the two summations and get
    \begin{align*}
        \mL_{\Psi\OD[\alpha]} \, A 
        &= \sum_{y\in \Z^2} \int_{\R}\dd s \sum_{x\in \Z^2} -\i[\, \alpha ^{-1} \,  \e^{\i s \mL_H} \, [\, h_x,\, \alpha \, \Psi_y\,],\, A\, ]\\
        &= \sum_{y\in \Z^2} [\, (\Psi_y)\OD[\alpha], \, A\, ]\, .
    \end{align*}
    Statement (iv) then follows directly by using the off-diagonal property of Lemma \ref{OD-property Observable}.
\end{proof}

\section{Main results}

Compared to the discussion in the introduction, we consider not only the response to an adiabatic change in the chemical potential on the right half-plane (i.e.\ to adding $\eps \La_1$ to the Hamiltonian), but also the response to adding generalized step function potentials $\eps \La_f$ (see Definition \ref{def:genswitch}). This shows, in particular, that the specific orientation of $\La_1$ and $\La_2$ relative to the lattice $\Z^2$   is irrelevant to the argument.

Given two generalized step functions $f$ and $g$, 
the additional   current flowing along the boundary of $\mathrm{supp}\, f$ into the region $\mathrm{supp}\, g$, in response to the perturbation $\eps \La_f$,   is  defined as the difference in expectation values of the rate of change $\i[H,\Lambda_g]$  of the charge in $\Lambda_g$ after and before the perturbation is turned on, i.e.\ by
\begin{equation}\label{def:deltagen}
 \Delta J_\eps = \w_\eps(\i[H,\Lambda_g])-  \w_0(\i[H ,\Lambda_g])\,.
\end{equation}

Our first theorem states that  $\Delta J_\eps$ is always well defined, nearly linear in the change $\eps$, and with linear coefficient given by the commutator formula for off-diagonal parts of the two generalized  step functions. Moreover, the linear coefficient depends only on $\w_0$, but not on $H$, $W$, and $S^\eps$.

The second theorem states that this linear coefficient depends solely on the relative orientation of the two generalized step functions, and that it can assume one of three values: zero, or plus or minus the Hall conductance. 

In summary, our results show that increasing the chemical potential in a  cone-like region $\mathrm{supp}(f)$  by $\eps$ for a system initially in a gapped ground state $\w_0$ induces an additional current that flows along the boundary of $\mathrm{supp}(f)$  and is nearly linear in $\eps$ and of the same magnitude (up to terms of order $\mathcal{O}(\eps^\infty)$) for all  regions $\mathrm{supp}(f)$ that contain a nonempty cone.
\vspace{-5mm}
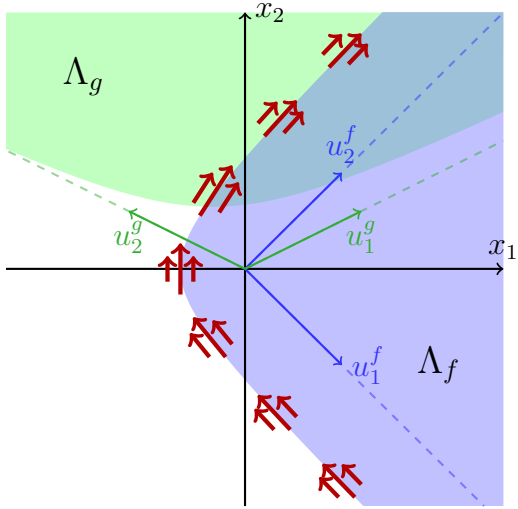
\begin{SCfigure}[2][ht]
    \label{fig:currentStepFunct}
\usetikzlibrary{patterns}

\begin{tikzpicture}[scale=0.85]
   \begin{scope}
   \clip
    plot[smooth,domain=-3.7:4,samples=200] (\x, {sqrt(1 
    +(\x+1)*\x / 4)}) -- (4,4) -- (-3.7,4) -- cycle;
    \fill[color=green!60,opacity=0.4] (-3.7,0) rectangle (4,4);
  \end{scope}
  \begin{scope}
    \clip 
    plot[smooth,domain=-1:3.5,samples=200] (\x, {sqrt((\x+2)*(\x+2)-1)})
      -- (4,4) -- (4,-4)
      -- plot[smooth,domain=3.5:-1,samples=200] (\x, {-sqrt((\x+2)*(\x+2)-1)})
      -- cycle;
    \fill[color=blue!60,opacity=0.4] (-3.7,-3.7) rectangle (4,4);
  \end{scope}
  \draw[->,thick] (-3.7,0) -- (4,0) node[above] {\large$x_1$};
  \draw[->,thick] (0,-3.7) -- (0,4) node[right] {\large$x_2$};

\draw[dashed,thick,color=blue!80,opacity=0.5] (0,0) -- (4,4); 
   \draw[->,thick,color=blue!80] (0,0) -- (1.5,1.5) node[above] {\large$u^f_2$};

    \draw[dashed,thick,color=blue!80,opacity=0.5] (0,0) -- (3.7,-3.7); 
    \draw[->,thick,color=blue!80] (0,0) -- (1.5,-1.5) node[right] {\large$u^f_1$};

     \draw[dashed,thick,color=ngreen!80,opacity=0.5] (0,0) -- (4,2);
    \draw[->,thick,color=ngreen!80] (0,0) -- (1.8,0.9) node[below] {\large$u^g_1$};

     \draw[dashed,thick,color=ngreen!80,opacity=0.5] (0,0) -- (-3.7,1.85);
    \draw[->,thick,color=ngreen!80] (0,0) -- (-1.8,0.9) node[below] {\large$u^g_2$};

 \draw[->,nred,ultra thick] (-1,-0.4) -- ++(0,0.8);
\draw[->,nred,ultra thick] (-1.2,-0.2) -- ++(0,0.4);
\draw[->,nred,ultra thick] (-.8,-0.2) -- ++(0,0.4);
  
  \foreach \xv in {1.3,2.3,3.3} {
    \pgfmathsetmacro\yv{sqrt(\xv*\xv - 1)}
    \pgfmathsetmacro\m{(\xv/\yv)}
    \draw[->,nred,ultra thick]
      (\xv-2,\yv) -- ++(0.5, {0.5*\m});
      \draw[->,nred,ultra thick]
      (\xv-2.1,\yv+0.2) -- ++(0.3, {0.3*\m});
      \draw[->,nred,ultra thick]
      (\xv-1.7,\yv+0.05) -- ++(0.3, {0.3*\m});
  }

  \foreach \xv in {1.7,2.7,3.7} {
    \pgfmathsetmacro\yv{sqrt(\xv*\xv - 1)}
    \pgfmathsetmacro\m{(\xv/\yv)}
    \draw[->,nred,ultra thick]
      (\xv-2,-\yv) -- ++(-0.5, {0.5*\m});
      \draw[->,nred,ultra thick]
      (\xv-2.25,-\yv) -- ++(-0.3, {0.3*\m});
      \draw[->,nred,ultra thick]
      (\xv-1.9,-\yv+0.2) -- ++(-0.3, {0.3*\m});
  }
    \node at (3,-1.5) {\Large$   \Lambda_f$};
    \node at (-2.5,3) {\Large$  \Lambda_g$};
\end{tikzpicture}
\caption{The quantity $\Delta J_\eps$ in \eqref{def:deltagen} expresses the rate of change of the charge in the region $\La_g$ in the NEASS state $\w_\eps$ that arises from an increase of the chemical potential in the region $\La_f$ by $\eps$ relative to the same quantity in the initial ground state $\w_0$. This rate of change   equals the current along the boundary of $\La_f$ and is the same for all generalized step functions $f$ (up to terms of order $\mathcal{O}(\eps^\infty)$).
Also the region $\La_g$ used to measure this current is determined by an arbitrary generalized step function $g$.
Note that     $\La_f$ and $\La_g$ are really  interactions, but we identify them for this heuristic illustration with the supports of the generalized step functions $ f$ and $ g$ defining  them. \vspace{10mm}
}
\vspace{-10mm}
\end{SCfigure}

\medskip

\begin{theorem}\label{thm:main}
    Let $f$ and $g$ be generalized step functions, such that the unit vectors $u_1^f, u_2^f, u_1^g, u_2^g$ are all different. Let $(\w_0, \,H,\, W,\, (S^\eps)_{\eps\in [-1,1]})$ be a gapped response system with respect to $f$. The expressions 
    \begin{align*}
        \Delta J_{\eps}  \coloneq \sum_{x\in \Z^2} \w_\eps(\i[\,H,\,\La_g\,]_x) - \w_0(\i[\,H,\,\La_g\,]_x)
    \end{align*}
    and
    \begin{align*}
        \w_0( \i[\, \La_g\OD,\, \La_f\OD\,]) \coloneq \sum_{x\in \Z^2} \sum_{y\in \Z^2} \w_0(\i[\,(\La_g\OD)_x,\, (\La_f\OD)_y\,])
    \end{align*}
    are absolutely convergent and it holds that 
    \begin{align*}
        \Delta J_{\eps}  =  \eps \, \w_0( \i\,[\, \La_g\OD,\, \La_f\OD\,]) + \mathcal{O}(\eps^\infty)\, .
    \end{align*}
    Furthermore, the value of $\w_0(\i \, [\, \La_g\OD,\, \La_f\OD])$ is the same for all gapped   systems with ground state~$\w_0$.
\end{theorem}
\begin{proof}
    The convergence of the two expressions follows from Lemma \ref{doublesum} and Lemma \ref{Rest sum}. To prove the main statement we first expand $\Delta J_{\eps}$ into a double sum by using Lemma \ref{OD-property Interaction} and Lemma \ref{lem: sum representation of generator}:
    \begin{align*}
        \Delta J_{\eps} &= \sum_{x\in \Z^2} -\w_\eps(\i\mL_{\La_g}\,h_x) + \w_0(\i\mL_{\La_g}\,h_x)\\
        &= \sum_{x\in \Z^2} -\w_\eps(\i\mL_{\La_g\ODeps}\,h_x) + \w_0(\i\mL_{\La_g\OD}\,h_x)\\
        &= \sum_{x\in \Z^2} \sum_{y\in \Z^2} -\w_\eps(\i \,[\,(\La_g\ODeps)_y,\,h_x\,]) +\w_0(\i \,[\,(\La_g\OD)_y, \,h_x\,]) \, .
    \end{align*}
    Lemma \ref{doublesum} tells us that this double sum converges absolutely and that we can thus change the order of the summations. This results in
    \begin{align*}
        \Delta J_{\eps} &= \sum_{y\in \Z^2} \w_\eps(\i \mL_H \, (\La_g\ODeps)_y) -\w_0(\i \mL_H\,(\La_g\OD)_y)\\
        &=\sum_{y\in \Z^2} \w_\eps(\i \mL_H \,(\La_g\ODeps)_y)\\
        &=\sum_{y\in \Z^2} \w_\eps(\i \mL_{H + \eps \La_f }\, (\La_g\ODeps)_y) - \sum_{y\in \Z^2} \eps \, \w_\eps(\i \mL_{\La_f \ODeps } \,(\La_g\ODeps)_y)\, .
    \end{align*}
    Lemma \ref{Rest sum} tells us that that the last step of splitting the expression into two sums is permitted since both of the sums converge absolutely. It further tells us that the first sum is $\mathcal{O}(\eps^\infty)$. We thus arrive at
    \begin{align*}
       \Delta J_{\eps} =  \eps \, \w_\eps( \i\,[\, \La_g\ODeps,\, \La_f\ODeps\,]) + \mathcal{O}(\eps^\infty)\, ,
    \end{align*}
    with 
    \begin{align*}
        \w_\eps( \i[\, \La_g\ODeps,\, \La_f\ODeps\,]) \coloneq \sum_{x\in \Z^2} \sum_{y\in \Z^2} \w_\eps(\i[\,(\La_g\ODeps)_x,\, (\La_f\ODeps)_y\,]) \, .
    \end{align*}
    
    Finally, we use the Chern-Simons Lemma \ref{CS-lemma} to arrive at the desired expression.

    For the last statement let $(\w_0, \,H_1,\, W_1)$ and $(\w_0, \,H_2,\, W_2 )$ be two gapped systems with the same ground state $\w_0$ and denote by $(\cdot)\OD[1]$ and $(\cdot)\OD[2]$ the respective off-diagonal mappings. Using Lemma \ref{lem: sum representation of generator} and Lemma \ref{OD-property Interaction}, it holds that
    \begin{align*}
        \w_0( \i \, [\, \La_g\OD[1] ,\, \La_f\OD[1] \,] ) &= \sum_{x\in \Z^2} \w_0(\i \mL_{\La_g\OD[1]} \, (\La_f\OD[1])_x)\\
        &= \sum_{x\in \Z^2} \w_0(\i \mL_{\La_g\OD[2]} \, (\La_f\OD[1])_x)\\
        &= \sum_{x\in \Z^2} \sum_{y\in \Z^2} \w_0(\i \, [ \, (\La_g\OD[2])_y , \, (\La_f\OD[1])_x \, ])\, .
    \end{align*}
    The sum is absolutely convergent due to Lemma \ref{lem: NEASS aut and OD estimates} and we can change the order of summation, resulting in
    \begin{align*}
        \w_0( \i \, [\, \La_g\OD[1] ,\, \La_f\OD[1] \,] ) &=  - \sum_{y\in \Z^2} \w_0(\i \mL_{\La_f\OD[1]}\, (\La_g\OD[2])_y) \\
        &=  - \sum_{y\in \Z^2} \w_0(\i \mL_{\La_f\OD[2]}\, (\La_g\OD[2])_y)\\
        &= \,\w_0( \i \, [\, \La_g\OD[2] ,\, \La_f\OD[2] \,] )\, . \qedhere
    \end{align*}
\end{proof}

\medskip

\begin{theorem}\label{thm:indep}
    Let $f$ and $g$ be generalized step functions, such that the unit vectors $u_1^f, u_2^f, u_1^g, u_2^g$ are all different and let $(\w_0, \,H,\, W)$ be a gapped system. The linear response coefficient $\w_0( \i\,[\, \La_g\OD,\, \La_f\OD\,])$ depends on the generalized step functions only through the order of the four unit vectors in the following way: 

    If the counter-clockwise order starting from $u_1^f$ is 
    $u_1^f$, $u_2^f$, $u_1^g$, $u_2^g$ or
    $u_1^f$, $u_1^g$, $u_2^g$, $u_2^f$ or
    $u_1^f$, $u_2^g$, $u_1^g$, $u_2^f$ or 
    $u_1^f$, $u_2^f$, $u_2^g$, $u_1^g$, then
    \[\w_0( \i\,[\, \La_g\OD,\, \La_f\OD\,]) \;=\;0 \, .\]

    If the counter-clockwise order starting from $u_1^f$ is $u_1^f$, $u_1^g$, $u_2^f$, $u_2^g$, then
    \[\w_0( \i\,[\, \La_g\OD,\, \La_f\OD\,]) \;=\;\w_0(\i[\Lambda_2\OD,\Lambda_1\OD]) \, .\]

    If the counter-clockwise order starting from $u_1^f$ is $u_1^f$, $u_2^g$, $u_2^f$, $u_1^g$, then 
    \[\w_0( \i\,[\, \La_g\OD,\, \La_f\OD\,]) \;=\;-\, \w_0(\i[\Lambda_2\OD,\Lambda_1\OD]) \, .\]
\end{theorem}
\begin{proof}
    We first assume that the ordering of the unit vectors is $u_1^f$, $u_2^f$, $u_1^g$, $u_2^g$ and note that 
    \begin{align*}
        \w_0( \i\,[\, \La_g\OD,\, \La_f\OD\,]) 
        &= \sum_{x\in \Z^2}\sum_{y\in \Z^2} \w_0(\i\, [\,(\La_g\OD)_x, \, (\La_f\OD)_y\,])
        \\
        &= \sum_{y\in \Z^2} \w_0(\i \, \mL_{\La_g}\,  (\La_f\OD)_y)
        \\
        &= \sum_{y\in \Z^2} \sum_{x\in \Z^2} \w_0(\i \, [\, g(x)\, n_x, \, (\La_f\OD)_y\,])\, .
    \end{align*}
    We want to show that the last sum is absolutely convergent. The function $g$ is supported on the set $C_\delta^g \cup C_1^g$, which is contained, up to a bounded region, in the cone with base point $0$ spanned by the vectors $ (u_1^{g,-\theta}, u_2^{g,\theta}) $, which are defined by rotating $u_1^g$ and $u_2^g$ by a small angle  counter-clockwise and clockwise respectively such that the resulting cone has a wider angle. By Lemma \ref{lem: half plane liouvillian}, $(\La_f\OD)_y$ is localized around the region $C_\delta^f$, which is, up a bounded region , contained in the cones with base point $0$ spanned by $(u_1^{f,-\theta}, u_1^{f,\theta})$ and by $(u_2^{f,-\theta}, u_2^{f,\theta})$, which are again defined by rotating by the angle $\theta$, creating two cones with angle $2\theta$ centered around $u_1^f$ and $u_2^f$ (see Figure~\ref{fig:conesProof} for an illustration). Hence, choosing $\theta$ small enough will result in three cones closures that only intersect in $0 \in \R^2$, allowing us to use Lemma \ref{lem:commutator bound} and Lemma \ref{lem: cone convegence} to conclude the absolute convergence. We can therefore change the order of summation and get
    \begin{align*}
        \w_0( \i\,[\, \La_g\OD,\, \La_f\OD\,]) 
        &= \sum_{x\in \Z^2}\sum_{y\in \Z^2} \w_0(\i \, [\, g(x)\, n_x, \, (\La_f\OD)_y\,])
        \\
        &= \sum_{x\in \Z^2} g(x)\, \w_0(-\i\,  \mL_{\La_f}\,  n_x ) = 0\, .
    \end{align*}

    \begin{center}

    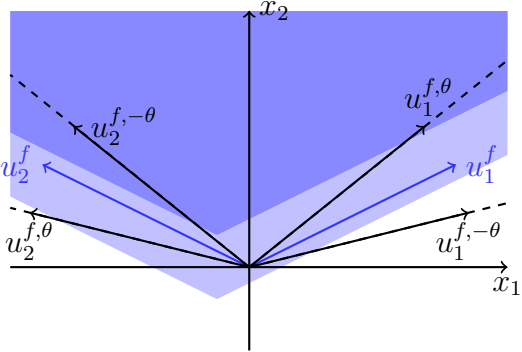
\begin{SCfigure}[2][ht]
        \label{fig:conesProof}
        \usetikzlibrary{patterns}
        \vspace{10mm}
        
        \begin{tikzpicture}[scale=0.85]
            \begin{scope}
                \clip
                plot[smooth,domain=-3.7:4,samples=200] (\x, {sqrt(0.00001 
                +(\x+.5)*(\x+.5)/4)-.5}) -- (4,4) -- (-3.7,4) -- cycle;
                \fill[color=blue!60,opacity=0.4] (-3.7,-2) rectangle (4,4);
            \end{scope}
            \begin{scope}
                \clip
                plot[smooth,domain=-3.7:4,samples=200] (\x, {sqrt(0.00001 
                +(\x+.5)*(\x+.5)/4)+.5}) -- (4,4) -- (-3.7,4) -- cycle;
                \fill[color=blue!80,opacity=0.4] (-3.7,-2) rectangle (4,4);
            \end{scope}
          
        
        
            \draw[->,thick] (-3.7,0) -- (4,0) node[below] {\large$x_1$};
            \draw[->,thick] (0,-1.3) -- (0,4) node[right] {\large$x_2$};

            \draw[->,thick,color=blue!80] (0,0) -- (-3.2,1.6)
            node[left] {\;\;\;\large$u^{f}_2$};
        
            \draw[dashed,thick] (0,0) -- (-3.7,0.925);
            \draw[->,thick] (0,0) -- (-3.4,0.85) node[below] {\large$u^{f,\theta}_2$};
            \draw[dashed,thick] (0,0) -- (-3.7,3);
            \draw[->,thick] (0,0) -- (-2.72,2.2) node[right] {\;\large$u^{f,-\theta}_2$};

            \draw[->,thick,color=blue!80] (0,0) -- (3.2,1.6)
            node[right] {\large$u^{f}_1$};;
        
            \draw[dashed,thick] (0,0) -- (4,1);
            \draw[->,thick] (0,0) -- (3.4,0.85) node[below] {\large$u^{f,-\theta}_1$};
            \draw[dashed,thick] (0,0) -- (4,3.23);
            \draw[->,thick] (0,0) -- (2.72,2.2) node[above] {\;\large$u^{f,\theta}_1$};
        
             
        \end{tikzpicture}
    
        \caption{
        For a generalized step function $f$ the figure shows $C^f_1$ as the upper dark blue shaded cone, $C^f_\delta$ as the light blue shaded strip, and $C^f_0$ as the white region. The vectors $u_1^f$ and $u_2^f$ span, up to a translation, the cone $C^f_1$. For any $\theta>0$, the set $C^f_\delta$ is contained in the union of the cones spanned by $u_1^{f,-\theta}$ and $u_1^{f, \theta}$ and by $u_2^{f, \theta}$ and $u_2^{f, -\theta}$, up to a bounded region. And the union of $C^f_1$ and $C^f_\delta$ is contained, up to a bounded region, in the cone spanned by $u_1^{f,-\theta}$ and $u_2^{f,\theta}$.
         \vspace{5mm}
        }
        \vspace{-5mm}
    \end{SCfigure}
    \end{center}  

    The other three orderings that result in zero can be reduced to this first one as follows: We observe that for the conjugate step function $\tilde{f}$  it holds that $u_1^{\tilde{f}}=u_2^{f}$ and $u_2^{\tilde{f}}=u_1^{f}$. So the three orderings correspond to the first one by either conjugating $f$ or $g$ or both. Additionally we observe that, due to gauge-invariance, we have
    \begin{align*}
        \w_0( \i\,[\, \La_g\OD,\, \La_{\tilde{f}}\OD\,]) 
        &= \sum_{x\in \Z^2} \w_0(-\i \, \mL_{\La_{\tilde{f}}}\, (\La_g\OD)_x)
        \\
        &= \sum_{x\in \Z^2} \w_0(-\i \, ( \mL_N - \mL_{\La_{f}}) \, (\La_g\OD)_x)
        \\
        &= \sum_{x\in \Z^2} \w_0(\i \, \mL_{\La_{f}} \, (\La_g\OD)_x)
        \\
        &= -\w_0( \i\,[\, \La_g\OD,\, \La_{f}\OD\,]) \, .
    \end{align*}
    This gives us the vanishing of the linear response coefficient for the claimed orderings.

    Let us now assume the ordering $u_1^f$, $u_1^g$, $u_2^f$, $u_2^g$ and let $g'$ be another generalized step function with the same ordering of unit vectors relative to $f$. For simplicity we assume to total ordering of the six unit vectors to be $u_1^f$, $u_1^g$, $u_1^{g'}$, $u_2^f$, $u_2^g$, $u_2^{g'}$ the other cases follow by the same argument. Our goal is to show that 
    \begin{align*}
        \w_0( \i\,[\, \La_g\OD,\, \La_{f}\OD\,]) 
        = \w_0( \i\,[\, \La_{g'}\OD,\, \La_{f}\OD\,])\, .
    \end{align*}
    To this end we look at the difference of the expressions:
    \begin{align*}
        \sum_{x\in \Z^2}\sum_{y\in \Z^2} \w_0(\i\, [\,(\La_g\OD)_x - (\La_{g'}\OD)_x, \, (\La_f\OD)_y\,])
        &= \sum_{y\in \Z^2} \w_0(\i \, ( \mL_{\La_g} - \mL_{\La_{g'}} )\,  (\La_f\OD)_y)
        \\
        &= \sum_{y\in \Z^2} \sum_{x\in \Z^2} \w_0(\i \, [\, ( g(x) - g'(x) )\, n_x, \, (\La_f\OD)_y\,])\, .
    \end{align*}
    As before we want to show the absolute convergence of the last sum.
    The expression $g(x) - g'(x)$ is supported in the union of the four sets  $ C_1^g \cap C_0^{g'} $, $ C_0^g \cap C_1^{g'} $, $C_\delta^g$, $C_\delta^{g'}$. This set lies, up to a bounded region, in the two cones with base point $0$ spanned by $ (u_1^{g,-\theta}, u_1^{g',\theta}) $ and by $ (u_2^{g,-\theta}, u_2^{g',\theta}) $, where again the additional superscript denotes a counter-clockwise rotation by the indicated angle. As before $(\La_f\OD)_y$ is localized around the region $C_\delta^f$, which is, up to a bounded region, contained in the cones with base point $0$ spanned by $(u_1^{f,-\theta}, u_1^{f,\theta})$ and by $(u_2^{f,-\theta}, u_2^{f,\theta})$. For $\theta$ small enough, this results in four cones with closures that intersect only in $0\in \R^2$, that support the two sets up to a bounded region. We again apply Lemma \ref{lem:commutator bound} and Lemma \ref{lem: cone convegence} to conclude the absolute convergence. This means we can switch the order of summation to get
    \begin{align*}
        \sum_{y\in \Z^2} \sum_{x\in \Z^2} \w_0(\i \, [\, ( g(x) - g'(x) )\, n_x, \, (\La_f\OD)_y\,])
        &=  \sum_{x\in \Z^2} ( g(x) - g'(x) ) \, \w_0(-\i \, \mL_{\Lambda_f} \,  n_x ) =0.
    \end{align*}
    Hence, we can replace $f$ and $g$ with any pair of generalized step functions, that have the same ordering of unit vectors without changing the value of the linear response coefficient. In particular we find
    \begin{align*}
        \w_0( \i\,[\, \La_g\OD,\, \La_f\OD\,]) \;=\;\w_0(\i[\Lambda_2\OD,\Lambda_1\OD]) \, .
    \end{align*}
    The remaining ordering, $u_1^f$, $u_2^g$, $u_2^f$, $u_1^g$, can be obtained from the one above by replacing $g$ with its conjugate $\tilde{g}$, which gives us
    \begin{align*}
        \w_0( \i\,[\, \La_g\OD,\, \La_{f}\OD\,])
        &= -\w_0( \i\,[\, \La_{\tilde{g}}\OD,\, \La_{f}\OD\,])
        \\
        &= -\w_0(\i[\Lambda_2\OD,\Lambda_1\OD])\, . \qedhere
    \end{align*}
\end{proof}

Next we observe that the linear response coefficients of Theorem \ref{thm:main} and in particular the Hall conductance,  of ground states connected by a locally generated automorphism are the same.

\medskip

\begin{theorem}\label{thm:const}
    Let $(\w_1,H_1,W_1)$ and $(\w_2,H_2,W_2)$ be gapped systems, $f$ and $g$ generalized step functions, such that the unit vectors $u_1^f, u_2^f, u_1^g, u_2^g$ are all different, and $\alpha$ a locally generated automorphism, such that $\w_2 = \w_1\circ\alpha$. Let $(\cdot)\OD[1]$ and $(\cdot)\OD[2]$ denote the respective off-diagonal maps. Then the systems have the same linear response coefficients:
    \begin{align*}
        \w_2(\i[\Lambda_g\OD[2],\Lambda_f\OD[2]] )= \w_1(\i[\Lambda_g\OD[1],\Lambda_f\OD[1]])\,.
    \end{align*}
\end{theorem}
\begin{proof}
    We observe that both $(\cdot)\OD[2]$ and $(\cdot)\OD[1\alpha]$ are off-diagonal maps for $\w_2$ (i.e. they satisfy property (iv) of Lemma \ref{OD-property Interaction}). Thus, the same argument as in the last part of the proof of Theorem \ref{thm:main}, yields
    \begin{align*}
        \w_2(\i[\Lambda_g\OD[2],\Lambda_f\OD[2]]) = \w_2(\i[\Lambda_g\OD[1\alpha],\Lambda_f\OD[1\alpha]]) = (\w_{1\alpha}(\i\,[\Lambda_g\OD[1\alpha],\Lambda_f\OD[1\alpha]]) \, .
    \end{align*}
    By the Chern-Simons lemma \ref{CS-lemma} the last expression is equal to $\w_1(\i[\Lambda_g\OD[1],\Lambda_f\OD[1]])$.
\end{proof}

The next theorem states that the constancy of the Hall conductance implies that the double commutator expression $\overline{\w_0}(\i[X_2\OD,X_1\OD])$ for the Hall conductivity involving modified position operators  is well defined  and agrees with the Hall conductance. This is known for non-interacting systems (e.g.\ \cite{elgart2005equality,marcelli2023localization}) and we show that it holds also for general gapped systems without any periodicity or 
homogeneity assumptions. In \cite{mtw2025} (see \cite{wmmmt2024} for periodic systems), we show that the linear response coefficient for the response of a gapped system to the application of a constant electric field is actually given by  $ \overline{\w_0}(\i[X_2\OD,X_1\OD])$.

\medskip

\begin{theorem}\label{thm:equiv}
    Let $(\w_0,H,W)$ be a gapped system. Then for any $x\in \Z^2$ the limit
    \begin{align*}
        \overline{\w_0}(\i[X_2\OD,X_1\OD]) \coloneq \lim_{k\to \infty} \frac{1}{\lvert B_k(x) \rvert} \sum_{y\in B_k(x)} \w_0(\i \mL_{X_2\OD} (X_1\OD)_y)
    \end{align*}
    exists and is equal to $\w_0( \i\,[\, \La_2\OD,\, \La_1\OD\,])$.
\end{theorem}
\begin{proof}
    Let $T$ be the family of standard translation automorphisms on $\mA$.
    We observe that due to \cite[Lemma E.4]{wmmmt2024} we have for $A \in D_\infty$
    \begin{align*}
        \mL_{X_1} \, A = \sum_{z\in \Z} \mL_{T_{(z,0)}\Lambda_1}\, A 
   \qquad\mbox{and}\qquad
        \mL_{X_2} \, A = \sum_{z\in \Z} \mL_{T_{(0,z)}\Lambda_2}\, A \, .
    \end{align*}
    This implies that
    \begin{align*}
        (X_1\OD)_y = \sum_{z\in \Z}   (T_{(z,0)}\Lambda_1\OD)_{y}
    \end{align*}
    and
    \begin{align*}
        \mL_{X_2\OD} A = \sum_{z\in \Z}  \mL_{(T_{(0,z)}\Lambda_2)\OD}\, A\,.
    \end{align*}
    Thus, we have for $x \in \Z^2$ and $k\in \N_0$
    \begin{align*}
        \sum_{y\in B_k(x)} \w_0(\i \mL_{X_2\OD} (X_1\OD)_y)
          \;= \sum_{y\in B_k(x)} \sum_{z\in\Z^2} \w_0(\i \mL_{T_z \Lambda_2} ((T_z\Lambda_1)\OD)_y)\, .
    \end{align*}
    We split the sum into two parts and add and subtract another sum, which results in
    \begin{align*}
        \sum_{y\in B_k(x)} \w_0(\i \mL_{X_2\OD} (X_1\OD)_y)
        & = \sum_{y\in B_k(x)} \sum_{z\in\Z^2} \w_0(\i \mL_{T_z \Lambda_2} ((T_z\Lambda_1)\OD)_y)\\
        & = \sum_{y\in \Z^2} \sum_{z\in B_k(x)} \w_0(\i \mL_{T_z \Lambda_2} ((T_z\Lambda_1)\OD)_y)\\
        & \quad - \sum_{y\in \Z^2\setminus B_k(x)} \sum_{z\in B_k(x)} \w_0(\i \mL_{T_z \Lambda_2} ((T_z\Lambda_1)\OD)_y)\\
        & \quad + \sum_{y\in B_k(x)} \sum_{z\in \Z^2 \setminus B_k(x)} \w_0(\i \mL_{T_z \Lambda_2} ((T_z\Lambda_1)\OD)_y)\, .
    \end{align*}
    The splitting and merging of sums is justified since we will now see that all sums involved are absolutely convergent.
    
    We see that by Theorem \ref{thm:indep} the first term is absolutely convergent and in fact equal to 
    \begin{align*}
        \lvert B_k(x) \rvert \,  \w_0( \i\,[\, \La_2\OD,\, \La_1\OD\,]) \ .
    \end{align*}
    It remains to show that the other two sums are each absolutely convergent with a bound that vanishes for large $k$ when compared to $\lvert B_k(x) \rvert$. We only consider the second sum, as the third can be treated analogously. Using Lemma \ref{lem: half plane liouvillian} and Lemma \ref{lem: NEASS aut and OD estimates} implies that there are constants $c_1$ and $c_2$, independent of $x,k,z,y$, such that
    \begin{align*}
        \sum_{y\in \Z^2\setminus B_k(x)} \sum_{z\in B_k(x)} \lVert  \mL_{T_z \Lambda_2} ((T_z\Lambda_1)\OD)_y) \rVert
        & \leq   \sum_{y\in \Z^2\setminus B_k(x)} \sum_{z\in B_k(x)} \frac{c_1\, \lVert (T_z\Lambda_1)\OD_y \rVert_{6,y}}{(1+ \lvert y_1-z_1\rvert)^6}
        \\
        & \leq \sum_{y\in \Z^2\setminus B_k(x)} \sum_{z\in B_k(x)} \frac{c_2}{(1+ \lvert y_1-z_1\rvert)^6 \, (1+ \lvert y_2-z_2\rvert)^6}
        \\
        & \leq \sum_{y\in \Z^2\setminus B_k(x)} \sum_{z\in B_k(x)} \frac{c_2}{(1+ \lVert y-z\rVert)^6 }
        \\
        & \leq c_2 \sum_{y\in \Z^2\setminus B_k(x)} \frac{1}{(1+\mathrm{d}(y,B_k(x)))^3} \sum_{z\in \Z^2} \frac{1}{(1+ \lVert z\rVert)^3 }
        \,.
    \end{align*}
    So we find that there is a constant $c_3$, independent of $k$, such that
    \begin{align*}\label{eq: linear upper bound}
        \sum_{y\in \Z^2\setminus B_k(x)} \sum_{z\in B_k(x)} \lVert  \mL_{T_z \Lambda_2} ((T_z\Lambda_1)\OD)_y) \rVert
        &\leq c_3 \sum_{y\in \Z^2\setminus B_k(x)} \frac{1}{(1+\mathrm{d}(y,B_k(x)))^3} \numberthis
        \, ,
    \end{align*}
    We now partition the $\Z^2$ into the sets with $\lVert y - x \rVert = l$ for $l\in \N_0$, resulting in
    \begin{align*}
        \sum_{y\in \Z^2\setminus B_k(x)} \frac{1}{(1+\mathrm{d}(y,B_k(x)))^3} 
        &= \sum_{l=k+1}^\infty \frac{8\,l}{(1+ l-k)^3}
        \\
        &= \sum_{l=1}^\infty \frac{8\,(l+k)}{(1+ l)^3}
        \\
        &= \sum_{l=1}^\infty \frac{8\,l}{(1+ l)^3} + k \sum_{l=1}^\infty \frac{8}{(1+ l)^3}\, .
    \end{align*}
    Thus the upper bound in \eqref{eq: linear upper bound} vanishes in the limit $k\to \infty$ when divided by $\lvert B_k(x) \rvert = (2\,k+1)^2$.
\end{proof}

The next and final theorem makes precise the claim that  in a gapped ground state no macroscopic currents flow over the boundary of a cone-like region.

\medskip

\begin{theorem}\label{thm:nomacro}
    Let $g$ be a generalized step function and $(\w_0, H, W)$ a gapped system. For each $z \in \Z^2$ it holds that
    \begin{align*}
        \lim_{k\to\infty} \, \frac{1}{k} \sum_{x\in B_k(z)} \w_0(\i \, [\, H,\, \Lambda_g\,]_x) = 0 \, .
    \end{align*}
\end{theorem}
\begin{proof}
    Let $k \in \N_0$. We use the off-diagonal property (Lemma \ref{OD-property Interaction}), insert a zero term, and expand the expression with Lemma \ref{lem: sum representation of generator} to get
    \begin{align*}
        \sum_{x\in B_k(z)} \w_0(\i \, [\, H,\, \Lambda_g\,]_x) 
        & = - \sum_{x\in B_k(z)} \w_0(\i \, \mL_{\Lambda_g\OD}\, h_x) 
        \\
        & = - \sum_{x\in B_k(z)} \w_0(\i \, \mL_{\Lambda_g\OD}\, h_x) 
        - \sum_{y \in B_k(z)} \w_0(\i \, \mL_{H}\, (\Lambda_g\OD)_y )
        \\
        & = - \sum_{x\in B_k(z)} \sum_{y\in \Z^2 } \w_0(\i \, [\, (\Lambda_g\OD)_y, \, h_x\, ]) 
        + \sum_{y \in B_k(z)}\sum_{x\in \Z^2 } \w_0(\i \, [\, (\Lambda_g\OD)_y, \, h_x\, ])
        \, .
    \end{align*}
    Both sums are absolutely convergent, allowing for them to partially cancel. This results in the sums
    \begin{align*}
        \sum_{x\in B_k(z)} \w_0(\i \, [\, H,\, \Lambda_g\,]_x) 
        & = - \sum_{x\in B_k(z)} \sum_{y\in \Z^2 \setminus B_k(z) } \w_0(\i \, [\, (\Lambda_g\OD)_y, \, h_x\, ])
        + \sum_{y \in B_k(z)}\sum_{x\in \Z^2 \setminus B_k(z) } \w_0(\i \, [\, (\Lambda_g\OD)_y, \, h_x\, ])
        \, , 
    \end{align*}
    which have an upper bound independent of of $k$, as we now show. We only consider the first sum, the second can be handled analogously. We use Lemma \ref{lem:commutator bound} and Lemma \ref{lem: NEASS aut and OD estimates} and see
    \begin{align*}
        \sum_{x\in B_k(z)} \sum_{y\in \Z^2 \setminus B_k(z) } \lvert \w_0(\i \, [\, (\Lambda_2\OD)_y, \, h_x\, ]) \rvert 
        & \leq \sum_{x\in B_k(z)} \sum_{y\in \Z^2 \setminus B_k(z) } \lVert   [\, (\Lambda_2\OD)_y, \, h_x\, ] \rVert
        \\
        & \leq \sum_{x\in B_k(z)} \sum_{y\in \Z^2 \setminus B_k(z) } 4^{3+7} \, \frac{(\lVert \Lambda_2\OD)_y \rVert_{7,y} \, \lVert h_x \rVert_{7,x}}{(1+ \lVert x- y \rVert)^7} 
        \\
        & \leq \sum_{x\in B_k(z)} \sum_{y\in \Z^2 \setminus B_k(z) }  \frac{c \, \lVert h_x \rVert_{7,x}}{(1+ \lVert x- y \rVert)^7 \, (1 + \mathrm{d}(y,C_\delta^g) )^6} 
        \, ,
    \end{align*}
    where $c$ is a constant independent of $x$ and $y$. We use that $\lVert h_x \rVert_{7,x} \leq 3\, \lVert H \rVert_7$ together with some basic estimates, which results in
    \begin{align*}
        \sum_{x\in B_k(z)} \sum_{y\in \Z^2 \setminus B_k(z) } \lvert \w_0(\i \, [\, (\Lambda_2\OD)_y, \, h_x\, ]) \rvert 
        & \leq \sum_{x\in B_k(z)} \sum_{y\in \Z^2 \setminus B_k(z) }  \frac{3 \,c\, \lVert H \rVert_{7}}{(1+ \lVert x- y \rVert)^4 \, (1 + \mathrm{d}(x,C_\delta^g) )^3\, (1 + \mathrm{d}(y,C_\delta^g) )^3}
        \\
        & \leq \sum_{x\in B_k(z)}  \frac{3 \,c\, \lVert H \rVert_{7}}{(1+ \mathrm{d}(x,\Z^2\setminus B_k(z)))^2 \, (1 + \mathrm{d}(x,C_\delta^g) )^3}
        \\
        &\qquad \times \sum_{y\in \Z^2 \setminus B_k(z) } \frac{1}{ (1+ \mathrm{d}(y, B_k(z)))^2 \,  (1 + \mathrm{d}(y,C_\delta^g) )^3}
        \, .
    \end{align*}
    For $n,l \in \N_0$ we now define the sets 
    \begin{align*}
        E^k_{l,n} 
        &\coloneq \{ x \in B_k(z) \,\, | \,\, l \leq \mathrm{d}(x,\Z^2\setminus B_k(z)) < l+1,  \,\, n \leq \mathrm{d}(x,C_\delta^g) < n+1 \}
        \\
        F^k_{l,n} 
        &\coloneq \{ y \in \Z^2 \setminus B_k(z) \,\, | \,\, l \leq \mathrm{d}(y, B_k(z)) < l+1,  \,\, n \leq \mathrm{d}(x,C_\delta^g) < n+1 \}
    \end{align*}
    and note that there is a constant $c_2 >0 $  such that  $\lvert E^k_{l,n} \rvert \leq c_2 \, (1+n) $ and $\lvert F^k_{l,n} \rvert \leq c_2 \, (1+n) $  for all $n, l\in\N_0$. This allows us to decompose the above sums as 
    \begin{align*}
        \sum_{x\in B_k(z)}  \frac{3 \,c\, \lVert H \rVert_{7}}{(1+ \mathrm{d}(x,\Z^2\setminus B_k(z)))^2 \, (1 + \mathrm{d}(x,C_\delta^g) )^3}
        &\leq \sum_{l=0}^k \sum_{n=0}^\infty \frac{3 \,c\, c_2 \, (1+n) \, \lVert H \rVert_{7}}{(1+ l)^2 \, (1 + n )^3}
    \end{align*}
    and 
    \begin{align*}
        \sum_{y\in \Z^2 \setminus B_k(z) } \frac{1}{ (1+ \mathrm{d}(y, B_k(z)))^2 \,  (1 + \mathrm{d}(y,C_\delta^g) )^3}
        &\leq \sum_{l=0}^\infty \sum_{n=0}^\infty \frac{c_2 \, (1+n) }{(1+ l)^2 \, (1 + n )^3}
        \, .
    \end{align*}
    Thus, we have shown that $\sum_{x\in B_k(z)} \w_0(\i \, [\, H,\, \Lambda_g\,]_x)$ has an upper bound independent of $k$, from which the claim follows.
\end{proof}

\medskip

\textbf{Acknowledgments}. We   thank Yoshiko Ogata for   helpful explanations and discussions, and for her hospitality during our visit to RIMS in Kyoto.
  We  also thank Guo Chuan Thiang for his useful comments, which prompted us to consider more general step functions.
This work was supported by the Deutsche Forschungsgemeinschaft (DFG, German Research Foundation) through TRR 352 (470903074) and FOR 5413 (465199066).

\appendix

\section{Locality of generalized step function operators}

\begin{definition}\label{def:genswitch}
    Let $C_1 \subset \R^2$ be an open cone, i.e.\  for some $x \in \R^2$ and a pair of unit vectors $(u_1,u_2)$ in $\R^2$ with $u_1 \neq u_2$ it holds that
    \begin{align*}
        C_1 = \{ y \in \R^2 \setminus \{x\} \ | \ 0 < \angle(u_1,y-x) < \angle(u_1,u_2) \}  \, ,
    \end{align*}
    where $\angle(x,y) \in [0,2\pi)$ is the angle from $x$ to $y$ measured counter-clockwise. Let further $C_0 \subset \R^2$ be such that for some $y \in \R^2$ 
    \begin{align*}
        C_0 =  y + \overline{C_1 }^c  
    \end{align*}    
    and $C_0 \cup C_1 = \emptyset$.
    Now let $f \colon \R^2 \to \R$ be a bounded function such that $f|_{C_0 } = 0$ and $f |_{C_1} = 1$.  We call such a function $f$ together with a choice of cones as above a \emph{generalized step function}. 

    Given a generalized step function $f$, we always use the following notation for the associated regions: 
    \begin{alignat*}{2}
        C^f_1 &\coloneq C_1  &&K^f_1 \coloneq \Z^2 \cap C_1\\
        C^f_0 &\coloneq C_0  &&K^f_0 \coloneq \Z^2 \cap C_0\\
        C^f_\delta &\coloneq (C_0 \cup C_1)^c \quad \quad \quad &&K^f_\delta \coloneq \Z^2 \cap C^f_\delta \, ,
    \end{alignat*}
    and we write $u_1^f$ and $u_2^f$ for the two unit vectors spanning the cone $C_1^f$.  Note that $ C^f_\delta \neq \emptyset$. 
    
  The interaction $\Lambda_f$ is defined as
    \[
        \La_f(\{x\}) := f(x) \,n_x\,\,\, \forall x\in \Z^2\quad\mbox{and}\quad \La_f(M) := 0 \quad\mbox{for all other sets $M \in P_0(\Z^2)$}\, .
    \]

    Finally we define the conjugated generalized step function of $f$, which we denote by $\tilde{f}$, as $\tilde{f} = 1-f$ with $C_1^{\tilde{f}} = C_0^f$ and $C_0^{\tilde{f}} = C_1^f$.
\end{definition}

\medskip
\begin{remark}\label{rem: half-plane step functions}
    Note that the half plane step functions, defined as
    \begin{align*}
        x \mapsto \begin{cases}
            1 & x_j \geq 0 \\
            0 & \text{else}
        \end{cases}
    \end{align*}
    for $j\in \{1,2 \}$ with the choice of cones $C_1 = \{ x \in \R^2\,\,| x_j >0 \}$ and $C_0 = \{ x \in \R^2\,\,| x_j <0 \}$, are generalized step functions and give rise to the interactions $\Lambda_j$.
\end{remark}

\medskip

\begin{lemma}\label{lem: cone convegence}
    Let $M_1, M_2 \subseteq \R^2$ and let $C_1, C_2, D_1, D_2$ be open cones such that their closures only intersect in the common base point $0 \in \R^2$. We assume that that $M_1$ and $M_2$ are contained in $C_1 \cup D_1$ and $C_2 \cup D_2$ respectively up to a bounded region. Then the following sums are absolutely convergent: 
    \begin{align*}\label{eq: doublesum}
        \sum_{x \in \Z^2} \sum_{y \in \Z^2} \frac{1}{(1+\mathrm{d}(x,M_1))^6}\, \frac{1}{(1+\mathrm{d}(y,M_2))^6} \, \frac{1}{(1+\lVert x-y \rVert)^6} \numberthis
        \, ,
    \end{align*}
    \begin{align*}\label{eq: singlesum}
        \sum_{x \in \Z^2} \frac{1}{(1+\mathrm{d}(x,M_1))^3}\, \frac{1}{(1+\mathrm{d}(x,M_2))^3} 
        \, .\numberthis
    \end{align*}
    And for each $m \in \N_0$ it holds that
    \begin{align*}\label{eq: sup}
        \sup_{x\in \Z^2 \setminus B_k(0)} \frac{1}{(1+\mathrm{d}(x,M_1))^m}\, \frac{1}{(1+\mathrm{d}(x,M_2))^m} 
        = \mathcal{O}(k^m) \numberthis
        \, ,
    \end{align*}
    where $\mathrm{d}$ denotes the distance with respect to the max-norm.
\end{lemma}
\begin{proof}
    We first estimate \eqref{eq: doublesum} by 
    \begin{align*}
        \hspace{2em}&\hspace{-2em}\sum_{x \in \Z^2} \sum_{y \in \Z^2} \frac{1}{(1+\mathrm{d}(y,M_1))^6}\, \frac{1}{(1+\mathrm{d}(x,M_2))^6} \, \frac{1}{(1+\lVert x-y \rVert)^6}
        \\
        &\leq \sum_{x \in \Z^2} \frac{1}{(1+\mathrm{d}(x,M_1))^3} \, \frac{1}{(1+\mathrm{d}(x,M_2))^3} 
        \sum_{y \in \Z^2} \frac{1}{(1+\mathrm{d}(y,M_1))^3}\,\frac{1}{(1+\mathrm{d}(y,M_2))^3} 
        \, .
    \end{align*}
    We note that this is simply the square of \eqref{eq: singlesum} and that its absolute convergence follows from \eqref{eq: sup}.
    Let $r \in \N_0$ be the such that $B_r(0)$, the box with side length $2\, r$ around $0$,  contains all elements in $M_1$ and $M_2$ that do not lie in the cones. This gives us for each $x \in \Z^2$ and $m \in \N_0$ the estimate 
    \begin{align*}
        \hspace{2em}&\hspace{-2em}\frac{1}{(1+\mathrm{d}(x,M_1))^m}\, \frac{1}{(1+\mathrm{d}(x,M_2))^m}
        \\
        &\leq \frac{1}{(1+\max(0,\mathrm{d}(x,C_1 \cup D_1)-r))^m}\, \frac{1}{(1+\max(0,\mathrm{d}(x,C_2 \cup D_2)-r))^m}
        \, .
    \end{align*}
    Since the four cones have closures that are disjoint up to the base point, we can widen each one to both sides by a small angle $\theta$ and get a new set of four open cones $C_1^\theta, C_2^\theta, D_1^\theta, D_2^\theta$ that are still disjoint. Let us define $K_1 \coloneq C_1 \cup D_1$, $K_2 \coloneq C_2 \cup D_2$ and 
    $K_1^\theta \coloneq C_1^\theta \cup D_1^\theta$, $K_2^\theta \coloneq C_2^\theta \cup D_2^\theta$. Let $k,m \in \N_0$ and $x \in \Z^2$ such that $\lVert x \rVert \geq k$. If $x \in K_1^\theta$ then $x \in \Z^2 \setminus K_2^\theta$ and we have
    \begin{align*}
        \frac{1}{(1+\mathrm{d}(x,M_1))^m}\, \frac{1}{(1+\mathrm{d}(x,M_2))^m} 
        & \leq \frac{1}{(1+\max(0,\mathrm{d}(x,K_1)-r))^m}\, \frac{1}{(1+\max(0,\mathrm{d}(x,K_2)-r))^m}
        \\
        &\leq \frac{1}{(1+\max(0,\mathrm{d}(x,K_2)-r))^m}
        \\
        &\leq \frac{1}{(1+\max(0,k \sin(\theta) - r))^m}
        = \mathcal{O}(k^m)
        \, .
    \end{align*}
    If $x \in \Z^2 \setminus  K_1^\theta$  we have
    \begin{align*}
        \frac{1}{(1+\mathrm{d}(x,M_1))^m}\, \frac{1}{(1+\mathrm{d}(x,M_2))^m}
        & \leq \frac{1}{(1+\max(0,\mathrm{d}(x,K_1)-r))^m}\, \frac{1}{(1+\max(0,\mathrm{d}(x,K_2)-r))^m}
        \\
        &\leq \frac{1}{(1+\max(0,\mathrm{d}(x,K_1)-r))^m}
        \\
        &\leq \frac{1}{(1+\max(0,k \sin(\theta) - r))^m}
        = \mathcal{O}(k^m)
        \, .
        \qedhere
    \end{align*}
\end{proof}

\medskip

\begin{lemma}\label{lem: half plane liouvillian}
    Let $\nu, \, m \in \N_0$ and let $f$  be a generalized step function. Then
    \begin{align*}
        \sup_{x\in \Z^2} \sup_{A\in D_\infty\setminus \{0\}} \frac{\lVert \mL_{\La_f} \, A \rVert_{\nu,x} }{\lVert A \rVert_{\nu + m +3, x}}   (1+ \mathrm{d}(x,C^f_\delta))^m < \infty 
        \, ,
    \end{align*}
    where $\mathrm{d}$ denotes the distance with respect to the max-norm.
\end{lemma}
\begin{proof}
    Let $A\in D_\infty$ and  $x\in K^f_0 \cup K^f_\delta$. We  use Lemma \ref{lem:commutator bound} to get
    \begin{align*}
        \lVert \mL_{\Lambda_f} \, A \rVert_{\nu,x} &\leq \sum_{y \in K^f_\delta\cup K^f_1}  \lVert [\, f(y) \, n_y , \, A\, ] \rVert_{\nu,x}
        \\
        &\leq \sum_{y \in K^f_\delta\cup K^f_1}
        4^{\nu+m+6} \, \frac{\Vert f(y)\,  n_y \rVert_{\nu+m+3,y} \, \lVert A \rVert_{\nu+m+3,x}}{(1+\lVert x -y \rVert)^{m+3}}
        \\
        &\leq \frac{1}{(1+ \mathrm{d}(x,C^f_\delta\cup C^f_1))^m} \sum_{y \in K^f_\delta\cup K^f_1} 4^{\nu+m+6} \, \frac{\Vert f(y)\,n_y \rVert_{\nu+m+3,y} \, \lVert A \rVert_{\nu+m+3,x}}{(1+\lVert x -y \rVert)^{3}}
        \\
        &= \frac{1}{(1+ \mathrm{d}(x,C^f_\delta))^m} \sum_{y \in K^f_\delta\cup K^f_1} 4^{\nu+m+6} \, \frac{\Vert f(y)\,n_y \rVert_{\nu+m+3,y} \, \lVert A \rVert_{\nu+m+3,x}}{(1+\lVert x -y \rVert)^{3}}\, .
    \end{align*}
    For $x \in K^f_1$ just note that one can repeat the argument for the conjugated generalized step function $\tilde f $ and that  $\mL_{\La_f} A = -\mL_{\La_{\tilde f}} A$, since $A$ is gauge invariant. 
\end{proof}

\medskip

\begin{lemma}\label{lem: NEASS aut and OD estimates}
    Let $f$ and $g$ be generalized step functions and let  $(\w_0,\,H,\,W,\,(S^\eps)_{\eps\in[-1,1]})$ be a gapped response system with respect to $f$.  Let $\nu,\,m\in \N_0$. It holds that
    \begin{align*}
        \sup_{\eps \in [-1,1]} \sup_{x\in\Z^2} \sup_{A\in D_\infty \setminus \{0\}} \frac{\lVert (\beta_\eps - 1)\, A \rVert_{\nu,x} \, (1+\mathrm{d}(x,C_\delta^f))^m}{\norm{A}_{\nu+3+m,x}} 
        &< \infty \, ,
    \end{align*}
    \begin{align*}
        \sup_{\eps\in[-1,1]} \sup_{x\in \Z^2} \lVert (\La_g\ODeps)_x \rVert_{\nu,x} \, (1+\mathrm{d}(x,C_\delta^g))^m < \infty 
    \end{align*}
    and
    \begin{align*}
        \sup_{\eps\in[-1,1]} \sup_{x\in \Z^2} \norm{\beta_\eps\,(\Lambda_g\ODeps)_x - (\Lambda_g\OD)_x}_{\nu,x} \, (1+\mathrm{d}(x,C_\delta^g))^m\, (1+\mathrm{d}(x,C_\delta^f))^m < \infty
        \, ,
    \end{align*}
    where $\mathrm{d}$ denotes the distance with respect to the max-norm.
\end{lemma}
\begin{proof}
    Let $\eps \in [-1,1]$, $x\in \Z^2$ and $A\in D_\infty$. For the first statement we write $\beta_\eps -1 $ as an integral using the generator $\i\mL_{S^\eps}$ and apply Lemma \ref{lem: cocycles} and Lemma \ref{lem: sum representation of generator} to get
    \begin{align*}
        \lVert (\beta_\eps - 1)\, A \rVert_{\nu,x}
        &\leq \int_0^1 \dd t \, \lVert \e^{\i t \mL_{S^\eps}} \, \mL_{S^\eps} \, A \rVert_{\nu,x}
        \\
        &\leq \int_0^1 \dd t \, b(t) \, \lVert \mL_{S^\eps} \, A \rVert_{\nu,x}
        \\
        &\leq \int_0^1 \dd t \, b(t) \sum_{y\in \Z^2} \lVert [\, S^\eps_y, \, A \, ] \rVert_{\nu,x}
        \, , 
    \end{align*}
    with a function $b$ that is independent of $\eps, x, A$ and grows at most polynomially in $t$.
    From here Lemma \ref{lem:commutator bound} and the properties of the NEASS generator ( Definition \ref{def:NEASS}) give us
    \begin{align*}    
        &\lVert (\beta_\eps - 1)\, A \rVert_{\nu,x}
        \leq \int_0^1 \dd t \, b(t) \sum_{y\in \Z^2} 4^{\nu + 6+m} \, \frac{\lVert S^\eps_y \rVert_{\nu+3+m,y} \, \lVert A \rVert_{\nu+3+m,x}}{(1+\lVert x-y \rVert)^{3+m}}
        \\
        &\leq \int_0^1 \dd t \, b(t) \sum_{y\in \Z^2} 4^{\nu + 6+m} \, \frac{\lVert A \rVert_{\nu+3+m,x}\, \sup_{\eps \in [-1,1]} \sup_{z\in \Z^2} \lVert S^\eps_z \rVert_{\nu+3+m,z} \, (1 + \mathrm{d}(z,C_\delta^f))^m  \, }{(1+ \mathrm{d}(y,C_\delta^f))^m\,(1+\lVert x-y \rVert)^{3+m}}\, ,
    \end{align*}
    from which the desired statement follows. For the second statement we use Lemma \ref{lem: cocycles} and the super-polynomial decay of $W$ and Lemma \ref{lem: half plane liouvillian} to get a three constants $c_1, c_2, c_3$ and an at most polynomially growing function $b_1$ all independent of $x$ and $\eps$, such that
    \begin{align*}
        \lVert (\La_g\ODeps)_x \rVert_{\nu,x}
        &= \lVert \beta_\eps^{-1} \int_{\R} \dd s \, W(s) \,  \e^{\i \mL_H} \, \beta_\eps \, \mL_{\La_g}\, \beta_\eps^{-1} \, h_x \rVert_{\nu,x} 
        \\
        &\leq c_1 \int_{\R} \dd s \, \lvert W(s) \, b_1(s)  \rvert \, \lVert  \mL_{\La_g}\, \beta_\eps^{-1} \, h_x \rVert_{\nu,x}
        \\
        &\leq c_1 \, c_2 \, (1 + \mathrm{d}(x, C_\delta^g))^m \, \lVert  \beta_\eps^{-1} \, h_x \rVert_{\nu+3+m,x}
        \\
        &\leq c_1 \, c_2 \, (1 + \mathrm{d}(x, C_\delta^g))^m \, c_3 \,  \lVert  h_x \rVert_{\nu+3+m,x}\, .
    \end{align*}
    We used the uniform boundedness of the NEASS generator (Definition \ref{def:NEASS}) to choose $b_1$ uniform in $\eps$. Hence, we have proven statement two.
    To treat the third statement we first assume $x \in C_0^g \cup C_\delta^g$ and insert the definition of the off-diagonal mapping. Additionally using  Lemma \ref{lem: cocycles} and the fact that $W$ decays faster that polynomially, results in a constant $C_1$, independent of $\eps$ and $x$, such that
    \begin{align*}
        \norm{\beta_\eps\,(\Lambda_g\ODeps)_x - (\Lambda_g\OD)_x}_{\nu,x}
        &\leq  \int_{\R} \dd s \sum_{y\in K_1^g\cup K_\delta^g}  \lvert W(s) \rvert \, \lVert \e^{\i s \mL_H} \, [\, (\beta_\eps -1) \, n_y, \,  h_x \,] \rVert_{\nu,x}
        \\
        &\leq C_1 \sum_{y\in K_1^g\cup K_\delta^g} \lVert [\, (\beta_\eps -1) \, n_y, \,  h_x \,] \rVert_{\nu,x}
        \, .
    \end{align*}
    With the commutator bound of Lemma \ref{lem:commutator bound} and the first statement, we arrive at
    \begin{align*}
        \norm{\beta_\eps\,(\Lambda_g\ODeps)_x - (\Lambda_g\OD)_x}_{\nu,x}
        &\leq C_1 \sum_{y\in K_1^g\cup K_\delta^g} 4^{\nu+2m+6} \, \frac{\lVert  (\beta_\eps -1) \, n_y\rVert_{\nu + 2m + 3,y} \, \lVert  h_x\rVert_{\nu + 2m + 3,x}}{(1+\lVert x- y\rVert)^{2m+3}}
        \\
        &\leq C_2 \sum_{y\in K_1^g\cup K_\delta^g}  \frac{\lVert  n_y\rVert_{\nu + 3m + 6,y} \, \lVert  h_x\rVert_{\nu + 2m + 3,x}}{(1+\mathrm{d}(y,C_\delta^f))^m \, (1+\lVert x- y\rVert)^{2m+3}}
        \,,
    \end{align*}
    where $C_2 $ is independent of $\eps$ and $x$.
    This can now be further estimated as
    \begin{align*}
        \norm{\beta_\eps\,(\Lambda_g\ODeps)_x - (\Lambda_g\OD)_x}_{\nu,x}
        &\leq  \frac{C_2}{(1+\mathrm{d}(x,C_\delta^f))^m \, (1+\mathrm{d}(x,C_\delta^g))^m} \sum_{y \in K_1^g\cup K_\delta^g}  \frac{\lVert  n_y\rVert_{\nu + 3m + 6,y} \, \lVert  h_x\rVert_{\nu + 2m + 3,x}}{(1+\lVert x- y\rVert)^{3}}
        \, .
    \end{align*}
    In the case where $x \in C_1^g$ one can do an analogous calculation with the conjugate step function $\tilde{g}$, where then $x\in C_0^{\title{g}}$, and use that due to gauge-invariance 
    \begin{equation*}
         \beta_\eps\,(\Lambda_g\ODeps)_x - (\Lambda_g\OD)_x = -(\beta_\eps\,(\Lambda_{\tilde{g}}\ODeps)_x - (\Lambda_{\tilde{g}}\OD)_x)
         \, .
         \qedhere
    \end{equation*}
\end{proof}

\medskip

\begin{lemma}\label{doublesum}
    Let $f$ and $g$ be generalized step functions such that the unit vectors $u_1^f, u_2^f, u_1^g, u_2^g$ are all different and let $(\w_0,\,H,\,W,\,(S^\eps)_{\eps\in[-1,1]})$ be a gapped response system with respect to $f$ and $\eps \in [-1,1]$. The sum 
    \begin{align*}
    \sum_{x\in \Z^2} \sum_{y\in \Z^2} -\w_\eps(\i [(\Lambda_g\ODeps)_y,h_x]) +\w_0(\i [(\Lambda_g\OD)_y, h_x])
    \end{align*}
    is absolutely convergent.
\end{lemma}
\begin{proof}
    Let $x,y \in Z^2$. We have 
    \begin{align*}
        |\w_\eps(\i [(\Lambda_g\ODeps)_y,h_x]) - \w_0(\i [(\Lambda_g\OD)_y, h_x])| &\leq \norm{[\beta_\eps(\Lambda_g\ODeps)_y,\beta_\eps h_x] - [(\Lambda_g\OD)_y, h_x]} \\
        &\leq \norm{[\,\beta_\eps\,(\Lambda_g\ODeps)_y - (\Lambda_g\OD)_y,\,\beta_\eps \, h_x\,]} + \norm{[\,(\Lambda_g\OD)_y, \, (\beta_\eps - 1) \, h_x\,]}
        \, .
    \end{align*}
    From here one can treat the two terms separately. For the first we use our bounds for the commutator and for $\beta_\eps\,\Lambda_g\ODeps - \Lambda_g\OD$ from Lemma \ref{lem:commutator bound} and Lemma \ref{lem: NEASS aut and OD estimates} as well as Lemma \ref{lem: cocycles}, resulting in
    \begin{align*}
        \norm{ [ \, \beta_\eps \, (\Lambda_g\ODeps)_y - (\Lambda_g\OD)_y , \, \beta_\eps \, h_x\,]} 
        &\leq \norm{[\,\beta_\eps\,(\Lambda_g\ODeps)_y - (\Lambda_g\OD)_y,\,\beta_\eps \, h_x\,]}_{0,x}
        \\
        &\leq 4^{3+12}\,  \frac{\norm{\beta_\eps\,(\Lambda_g\ODeps)_y- (\Lambda_g\OD)_y}_{12,y} \, \norm{\beta_\eps \, h_x}_{12,x}}{ (1+\norm{x-y})^{12} }
        \\
        &\leq \frac{c}{(1+\mathrm{d}(y,C_\delta^f))^6\, (1+\mathrm{d}(y,C_\delta^g))^6 \, (1+\norm{x-y})^{12} }
        \\
        &\leq \frac{c}{(1+\mathrm{d}(y,C_\delta^f))^6\, (1+\mathrm{d}(x,C_\delta^g))^6 \, (1+\norm{x-y})^6 }
        \, ,
    \end{align*}
    with a constant $c$ that is independent of $x$ and $y$. This expression is summable by Lemma \ref{lem: cone convegence}. For the Second term we similarly use our bounds for the commutator, $\La_g\OD$, and $(\beta_\eps -1 )$ from Lemma \ref{lem:commutator bound} and Lemma \ref{lem: NEASS aut and OD estimates}, to get
    \begin{align*}
        \norm{[\,(\Lambda_g\OD)_y, \, (\beta_\eps - 1) \, h_x\,]} 
        &\leq \norm{[\,(\Lambda_g\OD)_y, \, (\beta_\eps - 1) \, h_x\,]}_{0,x}
        \\
        & \leq 4^{3+6}\, \frac{\norm{(\Lambda_g\OD)_y}_{6,y} \, \norm{(\beta_\eps - 1) \,h_x}_{6,x}}{(1+\norm{x-y})^{6}}
        \\
        & \leq  \frac{c_2}{(1+\mathrm{d}(y,C_\delta^g))^6\, (1+\mathrm{d}(x,C_\delta^f))^6\,(1+\norm{x-y})^6}
        \, ,
    \end{align*}
    where $c_2$ is independent of $x$ and $y$, thus giving us a summable upper bound for both terms.
\end{proof}

\medskip

\begin{lemma}\label{Rest sum}
    Let $f$ and $g$ be generalized step functions such that the unit vectors $u_1^f, u_2^f, u_1^g, u_2^g$ are all different and let $(\w_0,\,H,\,W,\,(S^\eps)_{\eps\in[-1,1]})$ be a gapped response system with respect to $f$. For all $\eps \in [-1,1]$ the sums
    \begin{align*}
        \sum_{x\in \Z^2} \sum_{y\in \Z^2} [\, (\Lambda_g \ODeps)_x, \,  (\Lambda_f\ODeps)_y) \, ]
        \quad \text{and} \quad 
        \sum_{y\in \Z^2} \w_\eps(\i \mL_{H + \eps \Lambda_f } (\Lambda_g\ODeps)_y)
    \end{align*}
    are absolutely convergent and 
    \begin{align*}
        \sum_{y\in \Z^2} \w_\eps(\i \mL_{H + \eps \Lambda_f } (\Lambda_g\ODeps)_y) = \mathcal{O}(\eps^\infty)\, .
    \end{align*}
\end{lemma}
\begin{proof}
    Let $x, y \in \Z^2$. By Lemma \ref{lem: NEASS aut and OD estimates}, it holds that
    \begin{align*}
        \norm{[\, (\Lambda_g \ODeps)_x, \,  (\Lambda_f\ODeps)_y) \, ]} 
        &\leq \norm{[\, (\Lambda_g \ODeps)_x, \,  (\Lambda_f\ODeps)_y) \, ]}_{0,y}
        \\
        &\leq 4^{3+6} \frac{\norm{(\Lambda_g \ODeps)_x}_{6,x} \, \norm{(\Lambda_f \ODeps)_y}_{6,y}}{(1+\norm{x-y})^6}
        \\
        &\leq \frac{c}{(1+\mathrm{d}(x,C_\delta^g))^6 \, (1 + \mathrm{d}(x,C_\delta^g))^6 \, (1+\norm{x-y})^6}
        \, ,
    \end{align*}
    where $c$ is an $x$ and $y$ independent constant. The first sum is therefore absolutely convergent by Lemma \ref{lem: cone convegence}. Together with Lemma \ref{doublesum} this also implies absolute convergence of the second sum, since we can estimate it by
    \begin{align*}
        \sum_{y\in \Z^2} \lvert \w_\eps(\i \mL_{H + \eps \Lambda_f } (\Lambda_g\ODeps)_y) \rvert 
        &\leq \sum_{y\in \Z^2} \lvert \sum_{x \in \Z^2}  \w_\eps( [\, h_x + \eps (\Lambda_f\ODeps)_x, \, (\Lambda_g\ODeps)_y \, ] ) \rvert 
        \\
        &\leq \sum_{y\in \Z^2} \sum_{x \in \Z^2} \eps \, \lVert   [\, (\Lambda_f\ODeps)_x, \, (\Lambda_g\ODeps)_y \, ] \rVert 
        \\
        & + \sum_{y\in \Z^2} \lvert \w_\eps( \mL_H\, (\Lambda_g\ODeps)_y )- \w_0(\mL_H\, (\Lambda_g\OD)_y )  \rvert
        \, .
    \end{align*}
    To show that the second sum is $\mathcal{O}(\eps^\infty)$ we observe that for $\eps= 0$ is it $0$ and for $\eps \neq 0$ we split it in two parts
    \begin{align*}
        \sum_{y\in \Z^2} \w_\eps(\i \mL_{H + \eps \Lambda_f } (\Lambda_g\ODeps)_y) 
        = \sum_{y\in B_{\lvert 1/\eps \rvert}(0)} \w_\eps(\i \mL_{H + \eps \Lambda_f } (\Lambda_g\ODeps)_y) 
        + \sum_{y\in B_{\lvert 1/\eps \rvert}(0)^\mathsf{c} } \w_\eps(\i \mL_{H + \eps \Lambda_f } (\Lambda_g\ODeps)_y)\, . 
    \end{align*}
    For the first part we simply use the uniform NEASS property (Definition \ref{def:NEASS}) together with the fact that $\norm{(\Lambda_g\ODeps)_y}_{6,y}$ is bounded uniformly in $y$  and $\eps$ (Lemma \ref{lem: NEASS aut and OD estimates}). This yields
    \begin{align*}
        \sum_{y\in B_{\lvert 1/\eps \rvert}(0)} |\w_\eps(\i \mL_{H + \eps \Lambda_f } (\Lambda_g\ODeps)_y)| 
        &\leq \sum_{y\in B_{\lvert 1/\eps \rvert}(0)} \sup_{x \in \Z^2} \sup_{A \in D_\infty\setminus\{0\}} \frac{|\w_\eps(\mL_{H+\eps \La_f}A)|}{\norm{A}_{6,x}} \norm{(\Lambda_g\ODeps)_y}_{6,y}\\
        &= (2 \, \lvert 1/\eps \rvert + 1)^2 \, \mathcal{O}(\eps^\infty)\\
        &= \mathcal{O}(\eps^\infty) \, .
    \end{align*}
    For the second part we first insert a term equal to $0$ and split the sum again:
    \begin{align*}
        \hspace{2em}&\hspace{-2em} \sum_{y\in B_{\lvert 1/\eps \rvert}(0)^\mathsf{c} } \w_\eps(\i \mL_{H + \eps \Lambda_f } (\Lambda_g\ODeps)_y)
        \\
        &= \sum_{y\in B_{\lvert 1/\eps \rvert}(0)^\mathsf{c} } \w_0((\beta_\eps \, \i \mL_{H + \eps \Lambda_f } - \beta_\eps \, \i \mL_H + \beta_\eps \, \i \mL_H - \i \mL_H)\, (\Lambda_g\ODeps)_y)
        \\
        &= \sum_{y\in B_{\lvert 1/\eps \rvert}(0)^\mathsf{c} } \w_0((\beta_\eps \, \i \mL_{H + \eps \Lambda_f } - \beta_\eps \, \i \mL_H)\, (\Lambda_g\ODeps)_y) 
        + \sum_{y\in B_{\lvert 1/\eps \rvert}(0)^\mathsf{c} } \w_0(( \beta_\eps -1) \, \i \mL_H \, (\Lambda_g\ODeps)_y)
        \, .
    \end{align*}
    Simply estimating with the operator norm of $\w_0$ and $\beta_\eps$ leads to
    \begin{align*}
        \sum_{y\in B_{\lvert 1/\eps \rvert}(0)^\mathsf{c} } |\w_0((\beta_\eps \, \i \mL_{H + \eps \Lambda_f } - \beta_\eps \, \i \mL_H)\, (\Lambda_g\ODeps)_y)| 
        &\leq \sum_{y\in B_{\lvert 1/\eps \rvert}(0)^\mathsf{c} } \norm{(\mL_{H + \eps \Lambda_f } - \mL_H)\, (\Lambda_g\ODeps)_y}_{0,y}
        \\
        &=\eps \sum_{y\in B_{\lvert 1/\eps \rvert}(0)^\mathsf{c} } \norm{\mL_{\Lambda_f } \, (\Lambda_g\ODeps)_y}_{0,y}
        \, .
    \end{align*}
    Using Lemma \ref{lem: half plane liouvillian} and Lemma \ref{lem: NEASS aut and OD estimates} for some arbitrary $m\in\N_0$ yields
    \begin{align*}
        \hspace{2em}&\hspace{-2em}\sum_{y\in B_{\lvert 1/\eps \rvert}(0)^\mathsf{c} } |\w_0((\beta_\eps \, \i \mL_{H + \eps \Lambda_f } - \beta_\eps \, \i \mL_H)\, (\Lambda_g\ODeps)_y)|
        \\
        &\leq c_1 \, \eps \sum_{y\in B_{\lvert 1/\eps \rvert}(0)^\mathsf{c} } \frac{1}{(1+\mathrm{d}(y,C_\delta^f))^{3+m}} \, \norm{ (\Lambda_g\ODeps)_y}_{6+m,y}
        \\
        &\leq c_2 \, \eps \sum_{y\in B_{\lvert 1/\eps \rvert}(0)^\mathsf{c} } \frac{1}{(1+\mathrm{d}(y,C_\delta^f))^{3+m} \, (1+\mathrm{d}(y,C_\delta^g))^{3+m}}
        \\
        &\leq c_2 \, \eps \sup_{y\in B_{\lvert 1/\eps \rvert}(0)^\mathsf{c}} \frac{1}{(1+\mathrm{d}(y,C_\delta^f))^{m} \, (1+\mathrm{d}(y,C_\delta^g))^{m}} \sum_{y\in \Z^2 } \frac{1}{(1+\mathrm{d}(y,C_\delta^f))^{3} \, (1+\mathrm{d}(y,C_\delta^g))^{3}}
        \, ,
    \end{align*}
    with $\eps$ independent constants $c_1$ and $c_2$. The sum converges absolutely, while the supremum is $\mathcal{O}(\eps^m)$ as shown in Lemma \ref{lem: cone convegence}. 
    We deal with the remaining term similarly with the additional use of Lemma \ref{lem: sum representation of generator}:
    \begin{align*}
        \sum_{y\in B_{\lvert 1/\eps \rvert}(0)^\mathsf{c} } |\w_0(( \beta_\eps -1) \, \i \mL_H \, (\Lambda_g\ODeps)_y)|
        &\leq \sum_{y\in B_{\lvert 1/\eps \rvert}(0)^\mathsf{c} } \lVert (\beta_\eps -1) \, \mL_H \, (\Lambda_g\ODeps)_y \rVert_{0,y}
        \\
        &\leq \sum_{y\in B_{\lvert 1/\eps \rvert}(0)^\mathsf{c} } \frac{C_1}{(1+\mathrm{d}(y,C_\delta^f))^{m+3}} \, \lVert \mL_H \, (\Lambda_g\ODeps)_y \rVert_{6+m,y}
        \\
        &\leq \sum_{y\in B_{\lvert 1/\eps \rvert}(0)^\mathsf{c} } \frac{C_2}{(1+\mathrm{d}(y,C_\delta^f))^{m+3}} \, \lVert  (\Lambda_g\ODeps)_y \rVert_{5+12+2m,y}
        \\
        &\leq \sum_{y\in B_{\lvert 1/\eps \rvert}(0)^\mathsf{c} } \frac{C_3}{(1+\mathrm{d}(y,C_\delta^f))^{m+3} \,(1+\mathrm{d}(y,C_\delta^g))^{m+3}}
        \\
        &=\; \mathcal{O}(\eps^m)
        \, ,
    \end{align*}
    again with $\eps$ independent constants $C_1, \,C_2,\, C_3$. 
    Since this holds for all $m \in \N_0$, the claim is proven. 
\end{proof}

\section{Chern-Simons lemma}

In this section we prove the Chern-Simons lemma, Lemma~\ref{CS-lemma}, that shows that the double commutator expression ${\w_0}(\i[\Lambda_g\OD,\Lambda_f\OD]) $ is invariant under locally generated automorphisms.
It was already established in \cite{wmmmt2024} for $\overline{\w_0}(\i[X_2\OD,X_1\OD])$   assuming periodicity.

\medskip

\begin{lemma}\label{lem: commutator with step function}
    Let $\Phi$ be a $B_\infty$-interaction, $f$ a generalized step function and $A\in D_\infty$. It holds that
    \begin{align*}
        \mL_{\i[\Phi,\La_f]} \, A = \sum_{x\in \Z^2}  f(x)\,  [\, \i\mL_\Phi \, n_x, \, A \,] \, .
    \end{align*}
\end{lemma}
\begin{proof}
    We use Lemma \ref{lem:commutator bound} twice:
    \begin{align*}
        \sum_{x\in \Z^2} \sum_{y\in \Z^2}  \lVert [\,[\, f(y)\,n_y, \,\Phi_x\,],\,A\,] \rVert 
        &\leq \sum_{x\in \Z^2} \sum_{y\in \Z^2}  4^{3+3}\, \lVert [\,f(y)\, n_y, \,\Phi_x\,] \rVert_{3,x} \, \frac{\lVert A \rVert_{3,0}}{(1+ \lVert x \rVert)^3}
        \\
        &\leq \sum_{x\in \Z^2} \sum_{y\in \Z^2} 4^{3+3} \, 4^{3+6}\,  \frac{\lVert f(y) \, n_y \rVert_{6,y} \, \lVert \Phi_x  \rVert_{6,x} \, \lVert A \rVert_{3,0}}{(1 + \lVert y-x \rVert)^3\,(1+ \lVert x \rVert)^3}
        \, .
    \end{align*}
    Combining Lemma \ref{lem: sum representation of generator} with the absolute convergence of the above sum and the fact that the quasi-local terms of $\i[\Phi,\La_f]$ are given by $ \i[\Phi,\La_f]_{x} = -\i\mL_{\La_f}\, \Phi_x$, we can conclude the statement by simply changing the order of summation.
\end{proof}

\medskip

\begin{lemma}\label{lem: local aut estimates}
    Let $(\w_0,\,H,\,W)$ be a gapped system, $f$ a generalized step function, and $(\alpha_{u,v})_{(u,v)\in I^2}$ a locally generated cocycle generated by $(\Phi^v)_{v\in I}$. Let further $\nu,m \in \N_0$, $u,v \in I$ and $x\in \Z^2$. There exist increasing functions $b_1$ and $b_2$ that grow at most polynomially and are independent of $u$, $v$, $x$, $(\alpha_{u,v})_{(u,v)\in I^2}$, such that
    \begin{align*}
        \lVert \alpha_{u,v} \, (\La_f \OD[\alpha_{u,v}] )_x \rVert_{\nu,x}  \leq   \frac{b_1 (\lvert v-u \rvert)}{(1+\mathrm{d}(x,C_\delta^f))^m}
    \end{align*}
    \begin{align*}
        \lVert \alpha_{u,v} \, (\i[\, \Phi^{v}, \,  \La_f \, ] \OD[\alpha_{u,v}] )_x \rVert_{\nu,x}  \leq   \frac{b_2 (\lvert v-u \rvert)}{(1+\mathrm{d}(x,C_\delta^f))^m}
        \, ,
    \end{align*}
    where $\mathrm{d}$ denotes the distance with respect to the max-norm. It further holds that
    \begin{align*}
        \partial_v \, \alpha_{u,v} \, (\La_f\OD[\alpha_{u,v}])_x = \alpha_{u,v} \, ( \i\, [\, \Phi^{v}, \,  \La_f \, ] \OD[\alpha_{u,v}] )_x \, .
    \end{align*}
\end{lemma}
\begin{proof}
    For the first bound we expand the definition and see that from
    \begin{align*}
        \alpha_{u,v} \, (\La_f\OD[\alpha_{u,v}])_x = \i \int \dd s \, W(s) \, \e^{\i s \mL_H} \, \alpha_{u,v}^{-1} \, \mL_{\Lambda_f} \, \alpha_{u,v} \,   h_x 
    \end{align*}
    one can use Lemma \ref{lem: cocycles} once for each automorphism and Lemma \ref{lem: half plane liouvillian} on the Liouvillian to get the desired estimate.
    Expanding the definition further, we see that
    \begin{align*}
        \alpha_{u,v} \, (\La_f\OD[\alpha_{u,v}])_x 
        = \i  \int \dd s \sum_{y\in K_1^f \cup  K_\delta^f} W(s) \, \e^{\i s \mL_H} \, f(y)\, [\, \alpha_{u,v}\, n_y, \, h_x \, ] 
        \, .
    \end{align*}
    We will show that the term-wise derivative has an absolutely summable/integrable upper bound that is uniform in $v$, giving us the second bound as well as the last part of the statement. Let $m \in \N_0$ and $x \in K_0^f \cup  K_\delta^f$, then one has
    \begin{align*}
        \hspace{2em}&\hspace{-2em}\int_{\R} \dd s \sum_{y\in K_1^f \cup  K_\delta^f} \sup_{v\in I} |f(y) \, W(s)| \, \lVert \e^{\i s \mL_H} \,   [\, \alpha_{u,v}\, \mL_{\Phi^v} \, n_y, \, h_x \, ] \rVert
        \\
        &= \int_{\R} \dd s \sum_{y\in K_1^f \cup  K_\delta^f} \sup_{v\in I}  |f(y) \, W(s)| \, \lVert [\, \alpha_{u,v}\, \mL_{\Phi^v} \, n_y, \, h_x \, ] \rVert
        \\ 
        &\leq  \lVert f \rVert_\infty  \sum_{y \in K_1^f \cup  K_\delta^f} \sup_{v\in I}   \lVert [\, \alpha_{u,v}\, \mL_{\Phi^v} \, n_y, \, h_x \, ] \rVert \, .
    \end{align*}
    Lemmas \ref{lem:commutator bound}, \ref{lem: cocycles} and  \ref{lem: sum representation of generator} now give us an increasing and at most polynomially growing function $b$ such that
    \begin{align*}    
        \hspace{2em}&\hspace{-2em} \int_{\R} \dd s \sum_{y\in K_1^f \cup  K_\delta^f} \sup_{v\in I} |f(y)\,W(s)| \, \lVert \e^{\i s \mL_H} \, [\, \alpha_{u,v}\, \mL_{\Phi^v} \, n_y, \, h_x \, ] \rVert
        \\
        \leq & \lVert f \rVert_\infty \sum_{y\in K_1^f \cup  K_\delta^f} \sup_{v\in I}  b(\lvert v-u\rvert) \, \lVert \Phi^v \rVert_{3+2(m+3)} \, \frac{\norm{n_y}_{5+2(m+3),y} \, \lVert h_x \rVert_{3, x}}{(1+\lVert x-y \rVert)^{m+3} } 
        \, .
    \end{align*}
    Since for all $\nu \in \N_0$ we have $\sup_{v \in I} \lVert \Phi^v \rVert_\nu < \infty $ we can choose $b$ independently of $v$. In total we get a new function $b_2$ with the same properties such that
    \begin{align*}    
        \int_{\R} \dd s \sum_{y\in K_1^f \cup  K_\delta^f} \sup_{v\in I} |f(y)\,W(s)| \, \lVert \e^{\i s \mL_H} \, [\, \alpha_{u,v}\, \mL_{\Phi^v} \, n_y, \, h_x \, ] \rVert
        &\leq \frac{b_2 (\lvert v-u \rvert)}{(1+\mathrm{d}(x,C_\delta^f))^m}
        \, .
    \end{align*}
    In the case where instead $x \in K_1^f $ use the gauge-invariance of $\alpha_{u,v} \,   h_x $ and that for the conjugated step function $\tilde{f}$ it holds that $\mL_{\Lambda_{\tilde{f}}} = \mL_N -\mL_{\Lambda_f} $ to get the same bound.
    By dominated convergence we can pull the derivative inside the sum/integral and with with Lemma \ref{lem: commutator with step function} we can conclude that
    \begin{align*}
        \partial_v \, \alpha_{u,v} \, (\La_j\OD[\alpha_{u,v}])_x &= \i  \int \dd s \sum_{y\in \Z^2} W(s) \, \e^{\i s \mL_H} \,  f(y) \,  [\, \alpha_{u,v}\, \i\mL_{\Phi^v} \, n_y, \, h_x \, ] \\
        &= \i  \int \dd s \, W(s) \, \e^{\i s \mL_H} \, \alpha_{u,v} \, \mL_{\i[\Phi^v, \La_j]} \, \alpha_{u,v}^{-1} \, h_x\\
        &= \alpha_{u,v}\, (\i[\, \Phi^v, \, \La_j\, ]\OD[\alpha_{u,v}])_x \, .\qedhere
    \end{align*}    
\end{proof}

\medskip

\begin{lemma}\label{CS-lemma}
    Let $(\w_0,\, H,\,W)$ a gapped system, $f$ and $g$ generalized step functions, such that the unit vectors $u_1^f, u_2^f, u_1^g, u_2^g$ are all different, and $\alpha$ a locally generated automorphism. It holds that 
    \begin{align*}
        \w_{\alpha}(\i[ \, \La_g\OD[\alpha], \, \La_f\OD[\alpha] \, ]) 
        = \w_{0}(\i[\, \La_g\OD, \, \La_f\OD \, ]) 
        \, ,
    \end{align*}
    where 
    \begin{align*}
        \w_\alpha( \i[\, \La_g\OD[\alpha],\, \La_f\OD[\alpha]\,]) \coloneq \sum_{x\in \Z^2} \sum_{y\in \Z^2} \w_\alpha(\i[\,(\La_g\OD[\alpha])_x,\, (\La_f\OD[\alpha])_y\,])\, .
    \end{align*}
\end{lemma}
\begin{proof}
    We first note that by the same argument as in Lemma \ref{Rest sum}, the expression is absolutely convergent.
    Let $u,v \in I$ and $\alpha= \alpha_{u,v}$, where  $(\alpha_{s,t})_{(s,t)\in I^2}$ is generated by $(\Phi^s)_{s\in I}$. 
    The statement trivially holds true for $v=u$. To show the expression holds in general, we simply differentiate the left-hand side with respect to $v$ and show that the result vanishes for all $v$. To show that 
    \begin{align*}
        \w_{\alpha_{u,v}}(\i[ \, \La_g\OD[\alpha_{u,v}], \, \La_f\OD[\alpha_{u,v}] \, ]) 
        = \sum_{x\in \Z^2} \sum_{y\in \Z^2} \w_0( \i [ \, \alpha_{u,v} \,  (\La_g\OD[\alpha_{u,v}])_x, \, \alpha_{u,v} \, (\La_f\OD[\alpha_{u,v}])_y  \, ])
    \end{align*}
    is differentiable in $v$, we show that the term-wise differential is locally uniformly summable. Due to Lemma \ref{lem: local aut estimates} we have
    \begin{align*}
        \hspace{1em}&\hspace{-1em}\partial_v \, \w_0( \i [ \, \alpha_{u,v} \, (\La_g\OD[\alpha_{u,v}])_x, \, \alpha_{u,v} \, (\La_f\OD[\alpha_{u,v}])_y  \, ])
        \\ 
        &= \w_0( \i [ \, \alpha_{u,v} \,  ( \i[ \, \Phi^v , \, \La_g \, ] \OD[\alpha_{u,v}])_x, \, \alpha_{u,v} \, (\La_f\OD[\alpha_{u,v}])_y  \, ]) 
        + \w_0( \i [ \, \alpha_{u,v} \,  (\La_g\OD[\alpha_{u,v}])_x, \, \alpha_{u,v} \, (\i[\, \Phi^v, \, \La_f \, ]\OD[\alpha_{u,v}])_y  \, ])
        \\
        &\eqcolon S_1(v,x,y) + S_2(v,x,y) 
        \, .
    \end{align*}
    We can treat each term separately. Applying Lemmas \ref{lem:commutator bound} and \ref{lem: local aut estimates} results in an increasing at most polynomially growing function $b$, such that the following bound holds:
    \begin{align*}
        \lvert S_1(v,x,y) \rvert 
        &\leq 4^{3+4} \, \frac{\lVert (\i[\, \Phi^{v}, \,  \La_g \, ] \OD[\alpha_{u,v}] )_x \rVert_{6,x} \, \lVert (\La_f \OD[\alpha_{u,v}] )_y \rVert_{6,y}}{(1+\norm{x-y})^6}
        \\
        &\leq   \frac{b( \lvert v-u \rvert)}{(1+ \mathrm{d}(x,C_\delta^g))^6\, (1+ \mathrm{d}(y,C_\delta^f))^6 \, (1+\norm{x-y})^6} 
        \, .
    \end{align*}
    This bound is  locally uniformly summable with respect to $v$, due Lemma \ref{lem: cone convegence}. The same argument can be made for $S_2(v,x,y)$ and therefore, the map $I \, \to \, \R , \, v \, \mapsto \, \w_{\alpha_{u,v}}(\i[ \, \La_1\OD[\alpha_{u,v}], \, \La_2\OD[\alpha_{u,v}] \, ])$ is differentiable with 
    \begin{align*}
        \partial_v \, \w_{\alpha_{u,v}}(\i[ \, \La_g\OD[\alpha_{u,v}], \, \La_f\OD[\alpha_{u,v}] \, ]) = \sum_{x\in  \Z^2} \sum_{y\in \Z^2} S_1(v,x,y) + \sum_{x\in  \Z^2} \sum_{y\in \Z^2} S_2(v,x,y) \, .
    \end{align*}
    With Lemma \ref{lem: sum representation of generator} and Lemma \ref{OD-property Interaction} one sees that 
    \begin{align*}
        \sum_{x\in  \Z^2} \sum_{y\in \Z^2} S_1(v,x,y) = \sum_{y\in  \Z^2} \w_{\alpha_{u,v}}(\i\mL_{\i[\Phi^v,\La_g]}\, (\La_f\OD[\alpha_{u,v}])_y) = \sum_{x\in \Z^2} \w_{\alpha_{u,v}}(\i\mL_{\La_f}\, \i\mL_{\La_g} \, \Phi^v_x)
    \end{align*}
    and 
    \begin{align*}
        \sum_{x\in  \Z^2} \sum_{y\in \Z^2} S_2(v,x,y) =  - \sum_{x\in \Z^2} \w_{\alpha_{u,v}}(\i\mL_{\i[\Phi^v,\La_f]}\, (\La_g\OD[\alpha_{u,v}])_x) = - \sum_{y\in \Z^2} \w_{\alpha_{u,v}}(\i\mL_{\La_g}\, \i\mL_{\La_f} \, \Phi^v_y)\, .
    \end{align*}
    Therefore, the derivative is zero everywhere and $I \, \to \, \R , \, v \, \mapsto \, \w_{\alpha_{u,v}}(\i[ \, \La_g\OD[\alpha_{u,v}], \, \La_f\OD[\alpha_{u,v}] \, ])$ is a constant function.
\end{proof}

\printbibliography

\end{document}